\documentclass[aps,pra,notitlepage,superscriptaddress,showpacs]{revtex4-1}
\pdfoutput=1 
\usepackage{amsmath}
\usepackage{amssymb}
\usepackage{graphicx}

\begin{document}
\title{Off-axis optical vortices using double-Raman singlet and doublet light-matter schemes}
\author{Hamid Reza Hamedi}
\email{hamid.hamedi@tfai.vu.lt}
\affiliation{Institute of Theoretical Physics and Astronomy, Vilnius University,
Saul\.etekio 3, Vilnius LT-10257, Lithuania}

\author{Julius Ruseckas}
\email{julius.ruseckas@tfai.vu.lt}
\affiliation{Institute of Theoretical Physics and Astronomy, Vilnius University,
Saul\.etekio 3, Vilnius LT-10257, Lithuania}

\author{Emmanuel Paspalakis}
\email{paspalak@upatras.gr }
\affiliation{Materials Science Department, School of Natural Sciences, University
of Patras, Patras 265 04, Greece}

\author{Gediminas Juzeli\=unas}
\email{gediminas.juzeliunas@tfai.vu.lt}
\affiliation{Institute of Theoretical Physics and Astronomy, Vilnius University,
Saul\.etekio 3, Vilnius LT-10257, Lithuania}

\begin{abstract}
We study the formation of off-axis optical vortices propagating inside
a double-Raman gain atomic medium. The atoms interact with two weak
probe fields as well as two strong pump beams which can carry orbital
angular momentum (OAM). We consider a situation when only one of the
strong pump lasers carries an OAM. A particular superposition of
probe fields coupled to the matter is shown to form specific optical
vortices with shifted axes. Such off-axis vortices can propagate
inside the medium with sub- or superluminal group velocity depending
on the value of the two-photon detuning. The superluminal optical vortices are associated with the amplification as the energy of pump fields is transferred to the probe fields. 
The position of the peripheral vortices can be manipulated
by the OAM and intensity of the pump fields. We show that the exchange
of optical vortices is possible between individual probe beams and
the pump fields when the amplitude of the second probe field is zero
at the beginning of the atomic cloud. The model is extended to a more
complex double Raman doublet interacting with four pump fields. 
In contrast to the double-Raman-singlet, now the generation of the off-axis
sub- or superluminal optical vortices is possible even for zero two-photon
detuning.
\end{abstract}
\maketitle

\section{Introduction}

Electromagnetically induced transparency (EIT) \cite{Harris-EIT-1997,Fleischhauer-RevModPhys-2005}
is an optical effect in which the susceptibility of a weak probe field
is modified for an optically thick three-level $\Lambda$ atomic system.
The medium becomes transparent for the probe transition by applying
a stronger control field driving another transition of the $\Lambda$ system. A number
of important phenomena \cite{PaspalakisPhysRevA2002CPM,PaspalakisPhysRevA2002multi,PaspalakisPRA2002,Ruseckas-PhysRevA-2007} rely on the EIT,
including adiabatons \cite{Grobe-PhysRevLett-1994,Fleischhauer-PhysRevA-1996,Wang-phys.rev.lett.2001},
matched pulses \cite{Harris-PhysRevLett-1994,Cerboneschi-PhysRevA-1995},
giant optical nonlinearities \cite{Harris-PhysRevLett-1990,Deng-PhysRevA-1998,Wang-phys.rev.lett.2001,Kang-PhysRevLett-2003,HamediPhysRevA.Kerr}
and optical solitons \cite{Wu-PhysRevLett-2004,Wu-OL-2004}. Due to
the EIT the light pulses can be slowed down \cite{Harris-EIT-1997,Lene1999slowlgiht,Lukin-RevModPhys-2003,Juzeliunas-PhysRevLett-2004,Fleischhauer-RevModPhys-2005,Ruseckas2011PRA,Bao2011PRA,Ruseckas-PhysRevA.87-2013,Lee2014}
and also stored in the atomic medium by switching off the controlled laser \cite{Phillips-PhysRevLett-2001,Chien-Nature-2001,Lukin-Nature-2001,Juzeliunas2002,Lukin-RevModPhys-2003}.
This can be used to store 
the quantum state of light to the matter and back
to the light leading to important applications in quantum technologies \cite{Lukin-RevModPhys-2003,Lijun-J.Opt-2017,Hsiao-PhysRevLett-2018}.

In parallel, there have been important efforts dedicated to
the superluminal propagation of light pulses in coherently driven atomic media \cite{Jiang2007,Mahmoudi2006,Shang2009,AlexanderJOptics2010}.
An anomalous dispersion leading to the superluminal propagation can be naturally obtained within the absorption
band medium \cite{Chu1982} or inside a tunnel barrier \cite{Steinberg1993}.
Yet such a superluminal propagation is hardly observed due to losses \cite{Chu1982}.
Some novel approaches have been proposed to utilize transparent spectral
regions for fast light \cite{Chiao1993,Dogariu2001,Glasser2012,Bianucci2008,Julius2014superluminal}.
Specifically, it was shown that a linear anomalous dispersion can be
created in a Raman gain doublet and therefore distorsionless pulse
propagation is possible \cite{Dogariu2001}.

Vortex beams of light \cite{Allen1999,Miles-physToday-2004,Babiker2018} representing an example of
the singular optics, are of the fundamental interest
and offer many applications \cite{Babiker-PhysRevLett1994,Molina2001,Pugatch-PhysRevLet-2004,Dutton-PhysRevLett-2004,Bishop2004,Ruseckas-PhysRevA-2007,Chen-PhysRevA-2008,Lembessis-PhysRevA-2010,Ruseckas2011,Ding-OL-2012,WalkerPhysRevLett2012,Ruseckas-PhysRevA.87-2013,Lembessis-PhysRevA.89-2014,RadwellPhysRevLet2015,SharmaPhysRevA2017,Hamedi2018OE,Hamedi-PhysRevA-2018,Hamedipra2019,Moretti-PhysRevA-2009,Filippo2015,HamidCPT2019,Yin2019FWM,Jing2020FWM}.
The optical vortices are described by a wave field whose phase advances
around the axis of the vortex, and the associated wavefront carries an orbital
angular momentum (OAM)  \cite{Allen1999,Miles-physToday-2004,Babiker2018}. The phase of the on-axis optical vortices
advances linearly and monotonically with the azimuthal coordinate
reaching a multiple of $2\pi$ after completing a closed circle around the beam axis.
When two twisted beams each carrying an optical vortex are superimposed,
the resulting beam contains new vortices depending on the charge of
each vortex component \cite{Ivan2003,Galvez2006,Baumann2009,Hamedipra2019}.
Besides the on-axis vortices, the off-axis
optical vortices can be formed for which the vortex core is not on the
beam axis but moves about it \cite{Alois2001,Basistiy2003,Lee2005,Baumann2009,Zhang2017}.

In this article we propose a scenario for formation of off-axis
optical vortices in a four-level atom-light coupling scheme. We consider
a double-Raman gain medium interacting with two weak probe fields,
as well as two stronger pump laser beams which can carry the OAM. A specific
combination of the probe fields is formed with a definite group velocity
determined by the two-photon detuning. Note that the individual probe beams
do not have a definite group velocity when propagating inside the
medium. If one of the pump laser beams carries an optical vortex,
the resulting superposition beam exhibits off-axis vortices propagating
inside the medium with a sub- or superluminal group velocity depending
on the two-photon detuning. The position of the peripheral vortices around
the center can be manipulated by the OAM and the intensity of the pump fields.
It is shown analytically and numerically that the OAM of the pump fields
can be transferred to the individual probe beams when the amplitude
of the second probe field is zero at the beginning of the medium.
We also extend the model to a more complex double Raman doublet scheme interacting with
four pump fields.

The paper is organized as follows. In the next Section we consider the light propagation using a double Raman scheme and show that off-axis optical vortices with either slow or fast light properties can be created. In Section III we extend the model by considering formation of the off-axis optical vortices occurring under the slow or fast light propagation in the double Raman doublet scheme. Our results are summarized in Section IV.

\section{The double-Raman scheme}

\begin{figure}
\includegraphics[width=0.4\textwidth]{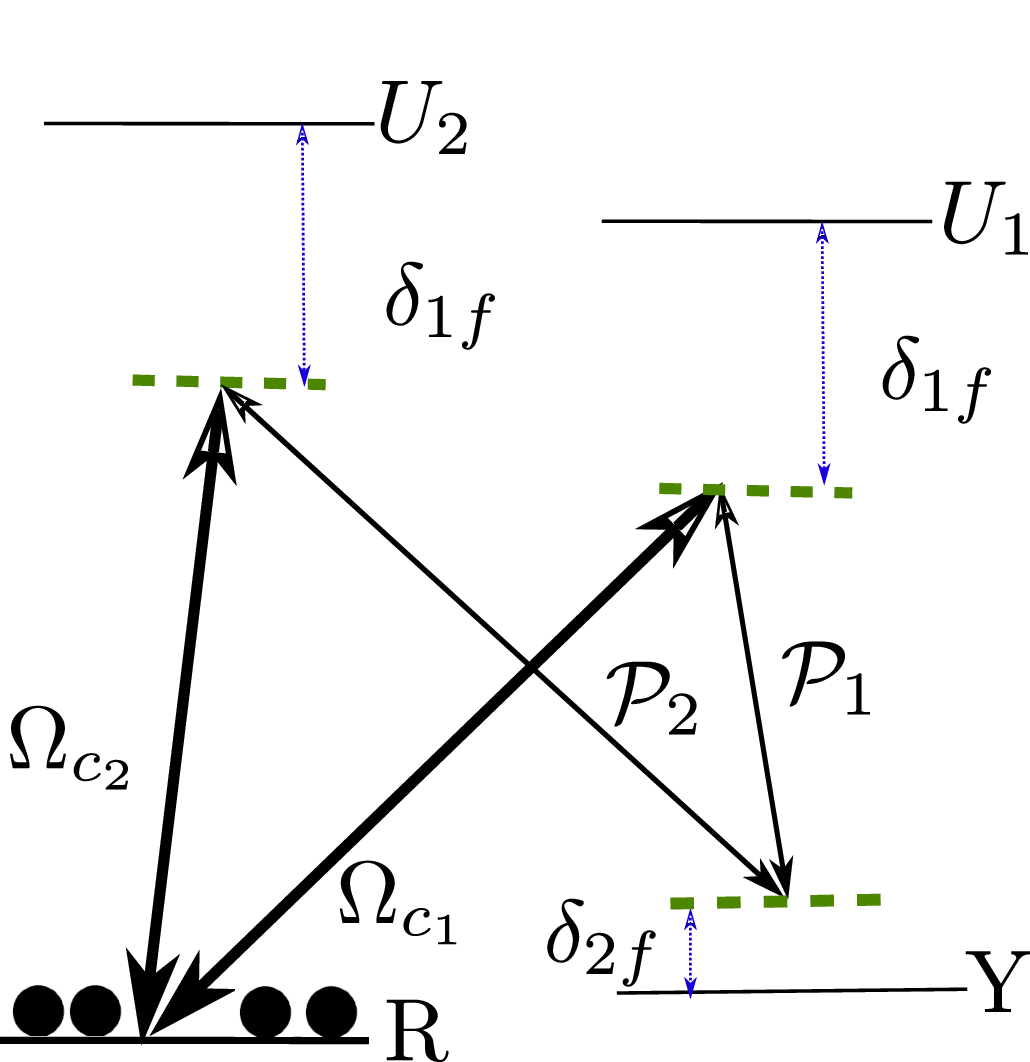}
\caption{Schematic diagram of the double Raman scheme}
\label{fig:DR}
\end{figure}
Let us consider propagation of two probe fields in an atomic medium with the Raman gain
described by the double Raman scheme illustrated in Fig.~\ref{fig:DR}. The atoms forming the medium
are characterized by two hyperfine ground
levels $R$ and $Y$ and two electronic excited levels $U_{1}$ and
$U_{2}$. The quantum state of the atoms is described by the probability amplitudes
$\textbf{\ensuremath{\Psi_{R}}}(\textbf{r},t)$, $\textbf{\ensuremath{\Psi_{Y}}}(\textbf{r},t)$,
$\textbf{\ensuremath{\Psi_{U_{1}}}}(\textbf{r},t)$, and $\textbf{\ensuremath{\Psi_{U_{2}}}}(\textbf{r},t)$
normalized to the atomic density $n$: $|\textbf{\ensuremath{\Psi_{R}}}|^{2}+|\textbf{\ensuremath{\Psi_{Y}}}|^{2}+|\textbf{\ensuremath{\Psi_{U_{1}}}}|^{2}+|\textbf{\ensuremath{\Psi_{U_{2}}}}|^{2}=n$.

The atoms interact with two weak probe fields with slowly varying
amplitudes $\mathcal{P}_{1}$ and $\mathcal{P}_{2}$, as well as two
strong pump lasers. The Rabi frequencies of the pump fields can be generally expressed as
\begin{equation}
\Omega_{c_{j}}=\mathcal{E}_{c_{j}}(r)e^{il_{j}\varphi},\label{eq:1}
\end{equation}
where
\begin{equation}
\mathcal{E}_{c_{j}}(r)=|\Omega_{c_{j}}|(r/w)^{|l_{j}|}e^{-r^{2}/w^{2}},\label{eq:2}
\end{equation}
is a fundamental Gaussian beam for $l_{j}=0$, while it describes
a Laguerre-Gaussian (LG) doughnut beam when $l_{j}\neq0$. Here
$\varphi$ is the azimuthal angle, $r$ describes the cylindrical
radius, $w$ denotes the beam waist parameter, and $|\Omega_{c_{j}}|$
($j=1,2$) is the strength of the pump beam.

The atoms are assumed to be initially in the ground level (Raman level)
$R$. The Rabi frequency and duration of the probe pulses are small
enough, so that the depletion of the ground level $R$ is neglected.
We work under the four-photon resonance condition $\omega_{p_{1}}-\omega_{c_{1}}=\omega_{p_{2}}-\omega_{c_{2}}$,
where $\omega_{p_{1}}$ and $\omega_{p_{2}}$ are the frequencies of the
probe beams, and $\omega_{c_{1}}$ and $\omega_{c_{2}}$ are the frequencies
of the pump beams.

After introducing the slowly varying atomic amplitudes we obtain the
following equations for slowly varying probe fields
\begin{align}
(\partial_{t}+c\partial_{z})\mathcal{P}_{1} & =i\alpha_{1}\Phi_{Y}^{*}\Phi_{U_{1}},\label{eq:3}\\
(\partial_{t}+c\partial_{z})\mathcal{P}_{2} & =i\alpha_{2}\Phi_{Y}^{*}\Phi_{U_{2}},\label{eq:4}
\end{align}
where $\alpha_{1}=\mu_{1}\sqrt{\omega_{p_{1}}/2\varepsilon_{0}\hbar}$,
$\alpha_{2}=\mu_{2}\sqrt{\omega_{p_{2}}/2\varepsilon_{0}\hbar}$ denote
the coupling  strength of the probe beams with the atoms, while $\mu_{1}$
and $\mu_{2}$ represent the dipole moments for the corresponding atomic
transitions. It should be noted that the diffraction terms containing
the transverse derivatives have been neglected in the Maxwell equations
(\ref{eq:3}) and (\ref{eq:4}). These terms are negligible if the
phase change of the probe fields due to these terms is much smaller
than $\pi$ \cite{Ruseckas-PhysRevA.87-2013,Hamedi-PhysRevA-2018,Hamedipra2019}.

Assuming that the strength of the coupling of the probe fields with the atoms
is the same $\alpha_{1}=\alpha_{2}=\alpha$, one arrives at the following equations for the
slowly varying atomic amplitudes
\begin{align}
i\partial_{t}\Phi_{U_{1}} & =\delta_{1f}\Phi_{U_{1}}-\alpha\mathcal{P}_{1}\Phi_{Y}-\mathcal{E}_{c_{1}}(r)e^{il_{1}\varphi}\Phi_{R},\label{eq:5}\\
i\partial_{t}\Phi_{U_{2}} & =\delta_{1f}\Phi_{U_{2}}-\alpha\mathcal{P}_{2}\Phi_{Y}-\mathcal{E}_{c_{2}}(r)e^{il_{2}\varphi}\Phi_{R},\label{eq:6}\\
i\partial_{t}\Phi_{Y} & =(\delta_{2f}-i\Gamma)\Phi_{Y}-\alpha\mathcal{P}_{1}^{*}\Phi_{U_{1}}-\alpha\mathcal{P}_{2}^{*}\Phi_{U_{2}},\label{eq:7}
\end{align}
where $\delta_{1f}=\omega_{U_{1}}-\omega_{R}-\omega_{c_{1}}=\omega_{U_{2}}-\omega_{R}-\omega_{c_{2}}$
describes the one-photon detuning, $\delta_{2f}=\omega_{p_{1}}-\omega_{c_{1}}+\omega_{Y}-\omega_{R}=\omega_{p_{2}}-\omega_{c_{2}}+\omega_{Y}-\omega_{R}$
represents the two-photon detuning and $\Gamma$ is the decay rate of
the level $Y$. Here, $\omega_{U_{1}}$, $\omega_{U_{2}}$ , and $\omega_{Y}$
are energies of the atomic states $U_{1}$, $U_{2}$ and $Y$, respectively.

We consider the case of monochromatic probe beams with the time-independent
amplitudes $\mathcal{P}_{1}$ and $\mathcal{P}_{2}$ and the spatially homogeneous
atomic amplitudes $\Phi_{R}$, $\Phi_{Y}$, $\Phi_{U_{1}}$, and $\Phi_{U_{2}}$.
We will look for the stationary solutions characterized by the time-independent atomic amplitudes $\Phi_{R}$, $\Phi_{Y}$,
$\Phi_{U_{1}}$, and $\Phi_{U_{2}}$, giving

\begin{align}
c\partial_{z}\mathcal{P}_{1}-i\alpha\Phi_{Y}^{*}\Phi_{U_{1}} & =0,\label{eq:8}\\
c\partial_{z}\mathcal{P}_{2}-i\alpha\Phi_{Y}^{*}\Phi_{U_{2}} & =0,\label{eq:9}\\
\delta_{1f}\Phi_{U_{1}}-\alpha\mathcal{P}_{1}\Phi_{Y}-\mathcal{E}_{c_{1}}(r)e^{il_{1}\varphi}\Phi_{R} & =0,\label{eq:10}\\
\delta_{1f}\Phi_{U_{2}}-\alpha\mathcal{P}_{2}\Phi_{Y}-\mathcal{E}_{c_{2}}(r)e^{il_{2}\varphi}\Phi_{R} & =0,\label{eq:11}\\
(\delta_{2f}-i\Gamma)\Phi_{Y}-\alpha\mathcal{P}_{1}^{*}\Phi_{U_{1}}-\alpha\mathcal{P}_{2}^{*}\Phi_{U_{2}} & =0.\label{eq:12}
\end{align}

For a large one-photon detuning $\delta_{1f}$ ($\delta_{1f}|\delta_{2f}-i\gamma|\gg\alpha^{2}|\mathcal{P}_{1,2}|^{2}$),
Eqs.~(\ref{eq:10}) and (\ref{eq:11}) give
\begin{align}
\Phi_{U_{1}} & =\frac{\mathcal{E}_{c_{1}}(r)}{\delta_{1f}}e^{il_{1}\varphi}\Phi_{R},\label{eq:13}\\
\Phi_{U_{2}} & =\frac{\mathcal{E}_{c_{2}}(r)}{\delta_{1f}}e^{il_{2}\varphi}\Phi_{R}.\label{eq:14}
\end{align}
Substituting Eqs.~(\ref{eq:13}) and (\ref{eq:14}) into Eq.~(\ref{eq:12})
yields
\begin{equation}
\Phi_{Y}=\frac{\alpha\Phi_{R}}{\delta_{1f}(\delta_{2f}-i\Gamma)}\left(\mathcal{E}_{c_{1}}(r)e^{il_{1}\varphi}\mathcal{P}_{1}^{*}+\mathcal{E}_{c_{2}}(r)e^{il_{2}\varphi}\mathcal{P}_{2}^{*}\right).\label{eq:15}
\end{equation}
Using Eqs.~(\ref{eq:13})-(\ref{eq:15}) the propagation
equation for both probe fields $P_{1}$ and $P_{2}$ (Eqs.~(\ref{eq:8})
and (\ref{eq:9})) take the form
\begin{align}
\partial_{z}P_{1}-i\beta\left(\frac{|\mathcal{E}_{c_{1}}(r)|^{2}\mathcal{P}_{1}+\mathcal{E}_{c_{1}}(r)\mathcal{E}_{c_{2}}^{*}(r)e^{i(l_{1}-l_{2})\varphi}\mathcal{P}_{2}}{(\delta_{2f}+i\Gamma)}\right) & =0,\label{eq:16}\\
\partial_{z}P_{2}-i\beta\left(\frac{\mathcal{E}_{c_{1}}^{*}(r)\mathcal{E}_{c_{2}}(r)e^{i(l_{2}-l_{1})\varphi}\mathcal{P}_{1}+|\mathcal{E}_{c_{2}}(r)|^{2}\mathcal{P}_{2}}{(\delta_{2f}+i\Gamma)}\right) & =0,\label{eq:17}
\end{align}
with
\begin{equation}
\beta=\frac{\alpha^{2}|\Phi_{R}|^{2}}{c\delta_{1f}^{2}}=\frac{\alpha^{2}n}{c\delta_{1f}^{2}}.\label{eq:18}
\end{equation}

We now introduce new fields representing superpositions of the original probe beams
\begin{equation}
\psi=\frac{1}{\mathcal{E}_{c}(r)}\left(\mathcal{E}_{c_{1}}^{*}(r)e^{-il_{1}\varphi}\mathcal{P}_{1}+\mathcal{E}_{c_{2}}^{*}(r)e^{-il_{2}\varphi}\mathcal{P}_{2}\right),\label{eq:19}
\end{equation}
\begin{equation}
\xi=\frac{1}{\mathcal{E}_{c}(r)}\left(\mathcal{E}_{c_{2}}(r)e^{il_{2}\varphi}\mathcal{P}_{1}-\mathcal{E}_{c_{1}}(r)e^{il_{1}\varphi}\mathcal{P}_{2}\right),\label{eq:20}
\end{equation}
where 
\begin{equation}
\mathcal{E}_{c}(r)=\sqrt{|\mathcal{E}_{c_{1}}(r)|^{2}+|\mathcal{E}_{c_{2}}(r)|^{2}}.\label{eq:21}
\end{equation}
is the total strength of the control fields. Calling on Eqs.~(\ref{eq:19}) and (\ref{eq:20}), one can rewrite
Eqs.~(\ref{eq:16}) and (\ref{eq:17}) as
\begin{equation}
\partial_{z}\psi-i\kappa\psi=0,\label{eq:22}
\end{equation}
\begin{equation}
\partial_{z}\xi=0,\label{eq:23}
\end{equation}
 where
\begin{equation}
\kappa=\beta\frac{\mathcal{E}_{c}^{2}(r)}{(\delta_{2f}+i\Gamma)}.\label{eq:24}
\end{equation}
 This behavior of the modes
$\psi$ and $\xi$ is similar to propagation in double-lambda system.
Eqs.~(\ref{eq:22}) and (\ref{eq:23}) clearly show that  
one of the superposition fields $\psi$ interacts with the atoms while
another field $\xi$ does not interact and propagates as in the free space.
The solution of Eq.~(\ref{eq:22}) reads
\begin{equation}
\psi(z)=\frac{1}{\mathcal{E}_{c}(r)}\left(\mathcal{E}_{c_{1}}^{*}(r)e^{-il_{1}\varphi}\mathcal{P}_{1}(0)+\mathcal{E}_{c_{2}}^{*}(r)e^{-il_{2}\varphi}\mathcal{P}_{2}(0)\right)e^{i\frac{\Gamma}{\delta_{1f}^{2}}\frac{\mathcal{E}_{c}^{2}(r)}{(\delta_{2f}+i\Gamma)}\frac{z}{L_{\Gamma}}},\label{eq:25}
\end{equation}
where
 
 \begin{equation}
\ensuremath{L_{\Gamma}=\frac{\Gamma c}{n\alpha^{2}}},\label{eq:LLL}
\end{equation}
determines the characteristic
length related to the decay of the excited level $Y$.

The group velocity of the light given by Eq.~(\ref{eq:25}) can be
calculated as
\begin{equation}
\nu_{g}=\frac{c}{1+\frac{\alpha^{2}n\mathcal{E}_{c}^{2}(r)}{\delta_{1f}^{2}}\frac{\Gamma^{2}-\delta_{2f}^{2}}{(\delta_{2f}^{2}+\Gamma^{2})^{2}}}.\label{eq:26-1}
\end{equation}
  Equation ~(\ref{eq:26-1}) is very similar to group velocity in a Raman system with
single probe beam.
Clearly, when $\Gamma<\delta_{2f}$ the group velocity exceeds $c$
providing the superluminality. On the other hand, the slow light propagates
in the medium when $\Gamma>\delta_{2f}$ ($\nu_{g}<c$ ). In particular, the superluminal propagation is associated with the amplification since the energy of pump fields is transferred to the probe fields. This can be
easily seen from the fact that the coefficient $\kappa$ in Eq.~(\ref{eq:24}) is a complex number.

\begin{figure}
\includegraphics[width=0.3\textwidth]{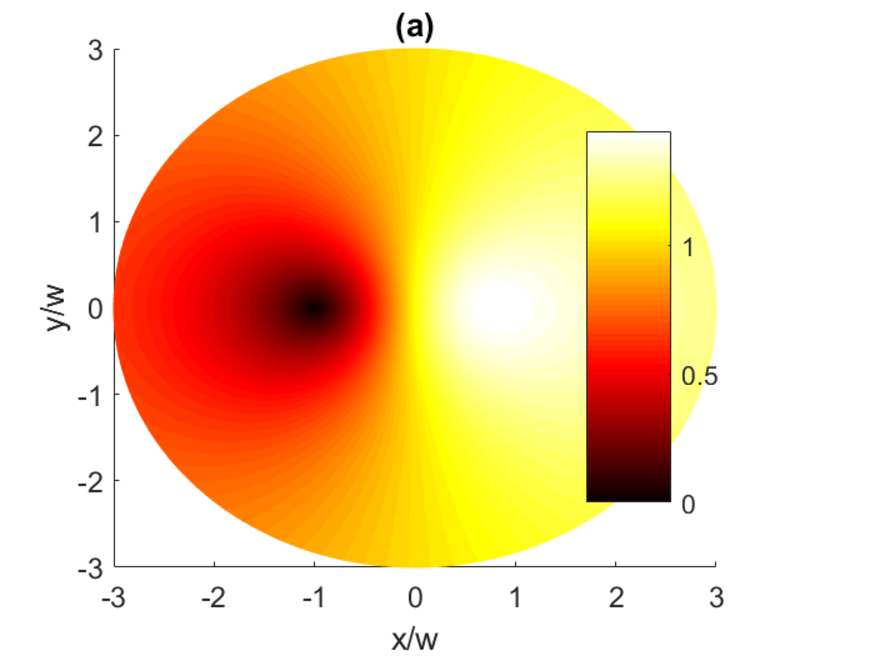}\includegraphics[width=0.3\textwidth]{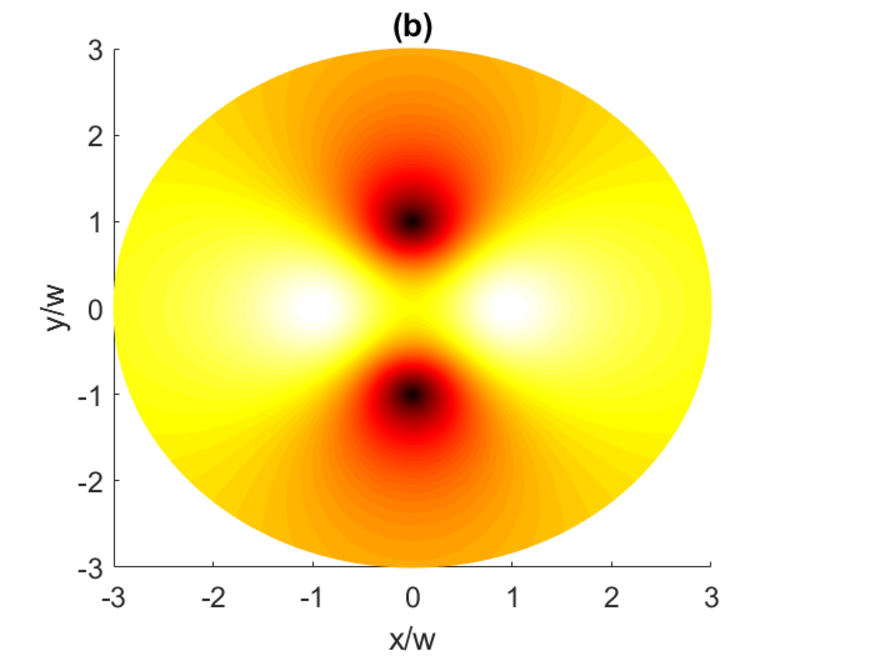}\includegraphics[width=0.3\textwidth]{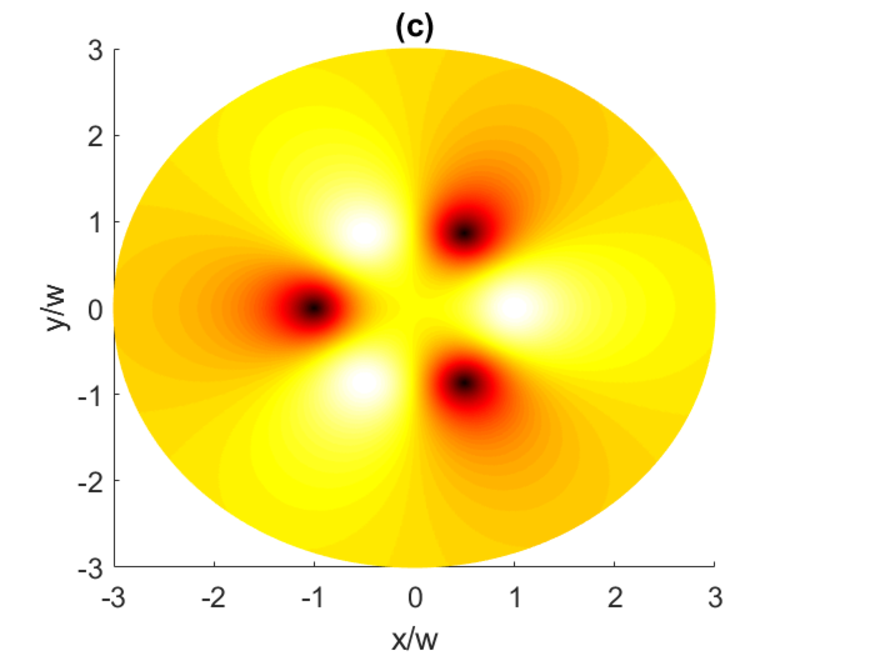}

\includegraphics[width=0.3\textwidth]{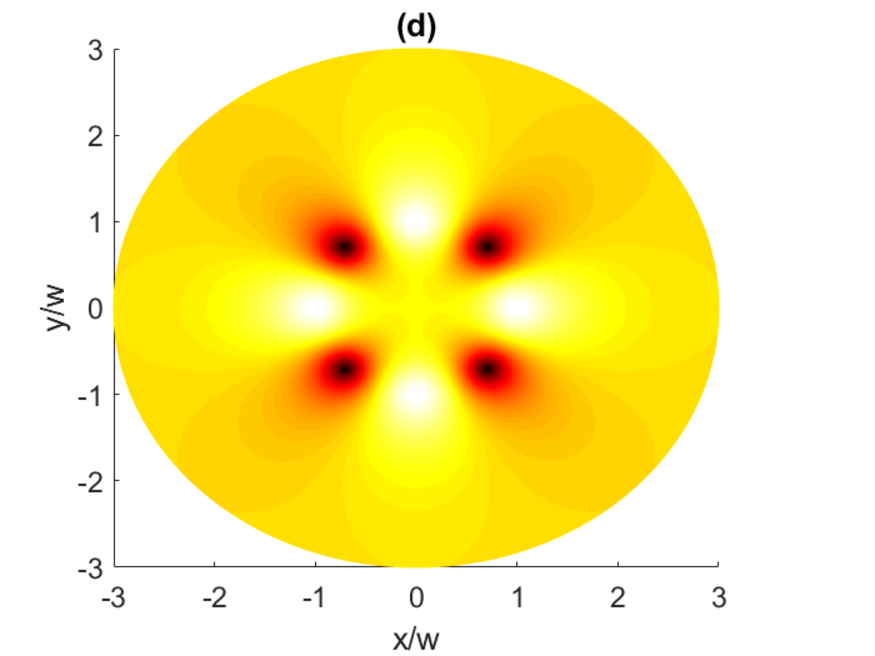}\includegraphics[width=0.3\textwidth]{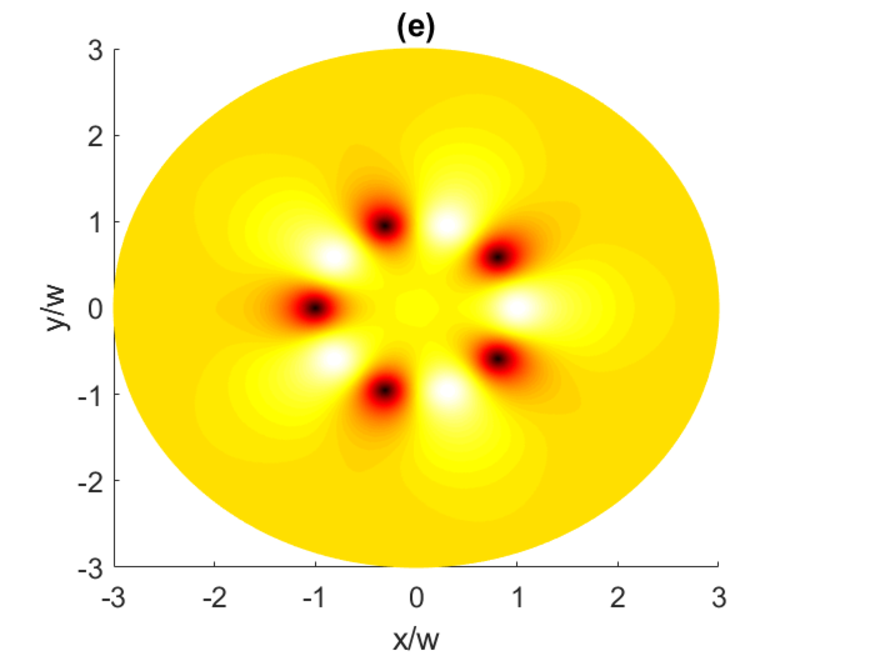}\includegraphics[width=0.3\textwidth]{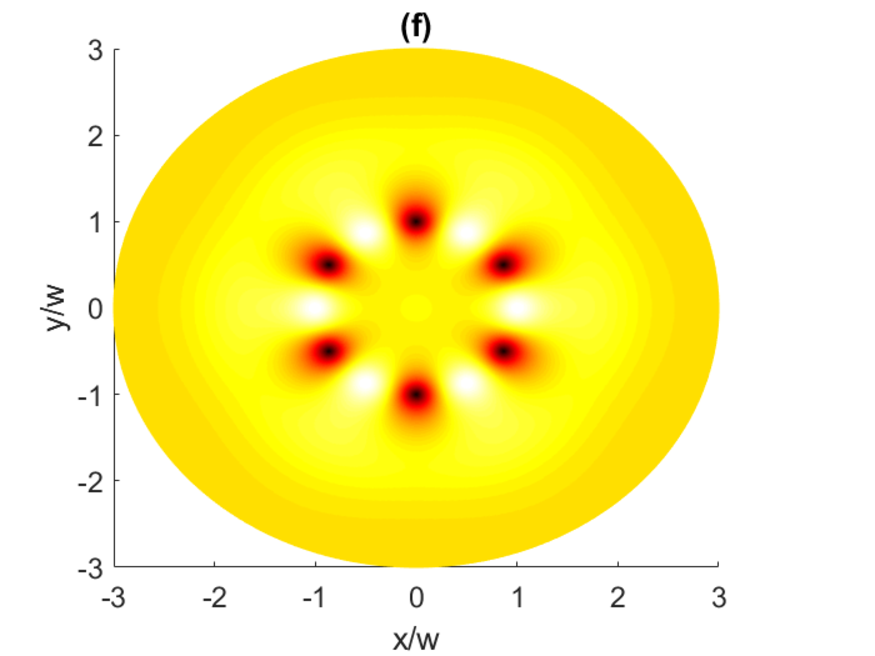}
\caption{Intensity distributions (in arbitrary units) of the superluminal superposition
beam $\psi$ featured in Eq.~(\ref{eq:25}) with different vorticities
$l_{1}=1-6$ (a)-(e). Here the parameters are $|\Omega_{c_{1}}|=|\Omega_{c_{2}}|=\Gamma$,
$\delta_{1f}=\Gamma$, $z=L_{\Gamma}$, $l_{2}=0$ and $\delta_{2f}=4\Gamma$.}
\label{fig:2}
\end{figure}

\begin{figure}
\includegraphics[width=0.3\textwidth]{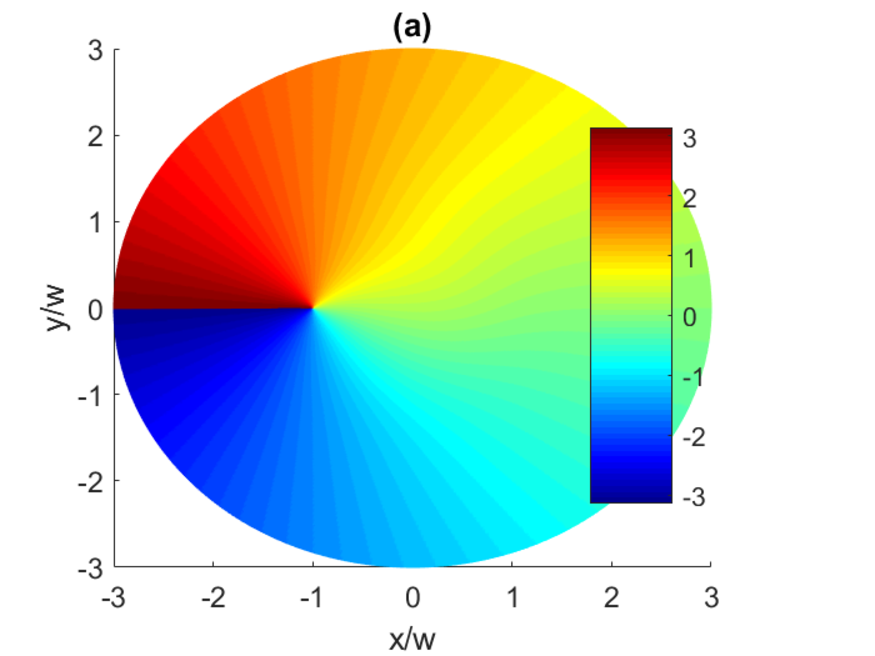}\includegraphics[width=0.3\textwidth]{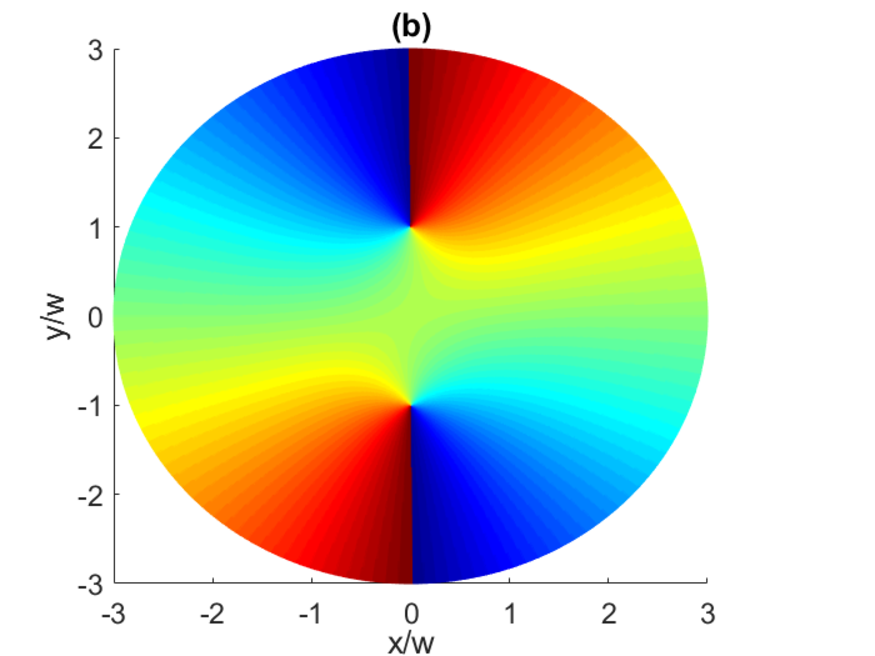}\includegraphics[width=0.3\textwidth]{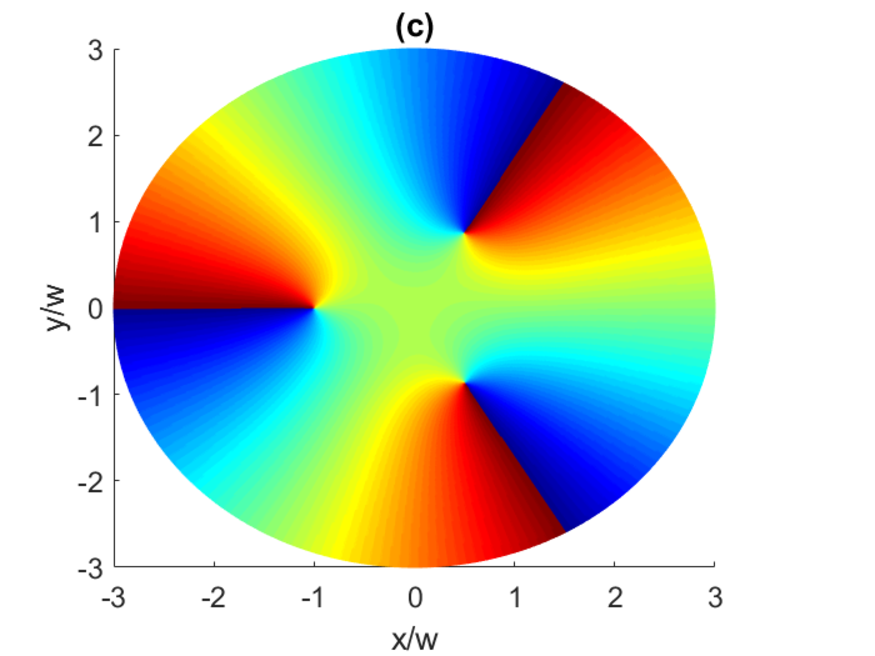}

\includegraphics[width=0.3\textwidth]{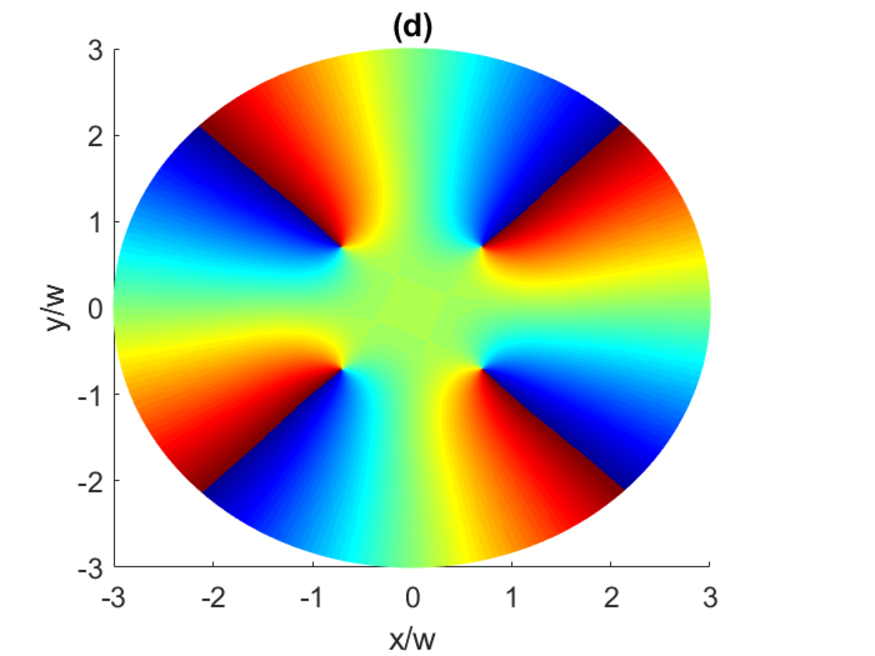}\includegraphics[width=0.3\textwidth]{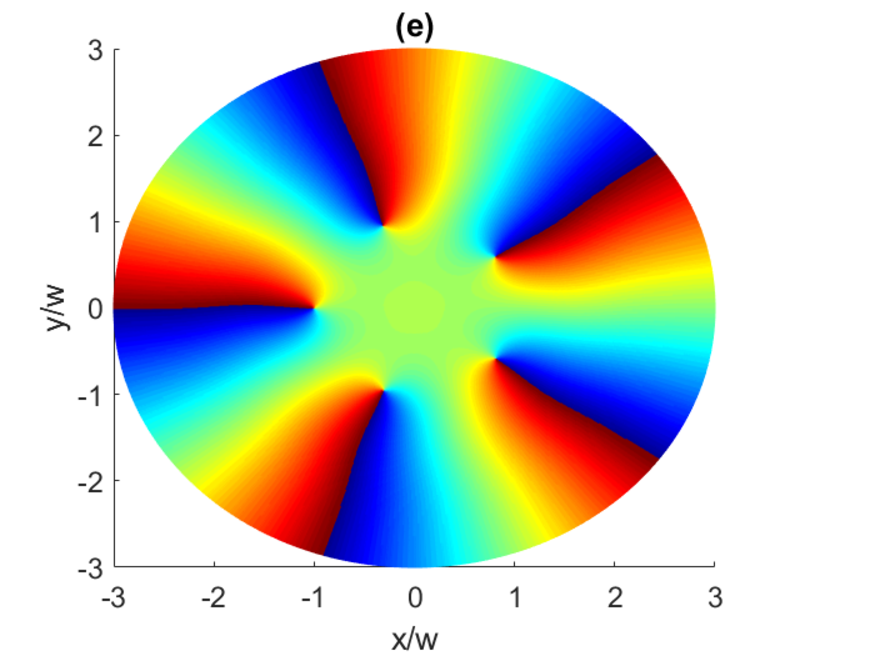}\includegraphics[width=0.3\textwidth]{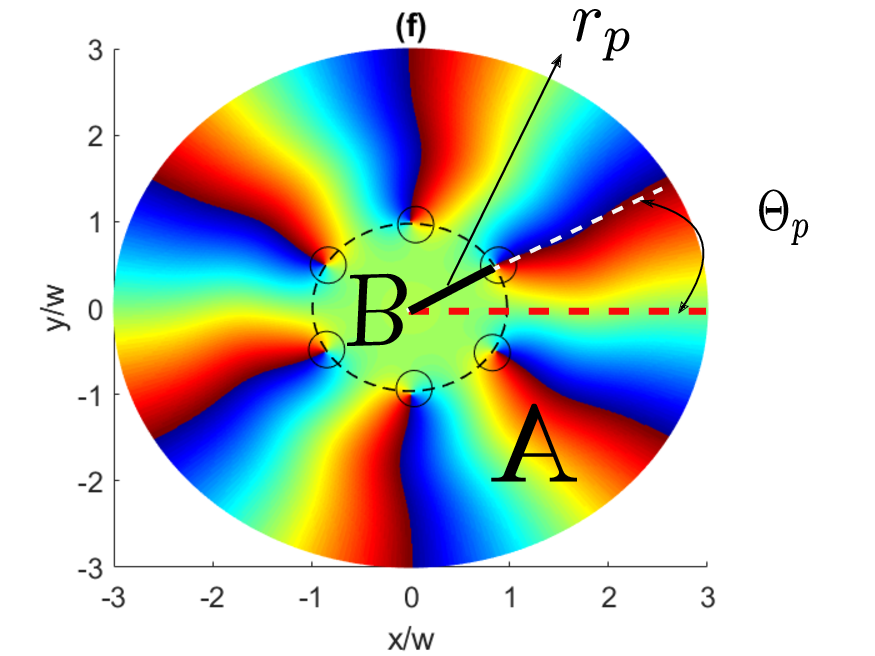}
\caption{The helical phase patterns of the superluminal superposition beam
$\psi$ featured in Eq.~(\ref{eq:25}) with different vorticities
$l_{1}=1-6$ (a)-(e). The parameters are taken to be the same as in Fig.~\ref{fig:2}. }
\label{fig:3}
\end{figure}

\begin{figure}
\includegraphics[width=0.3\textwidth]{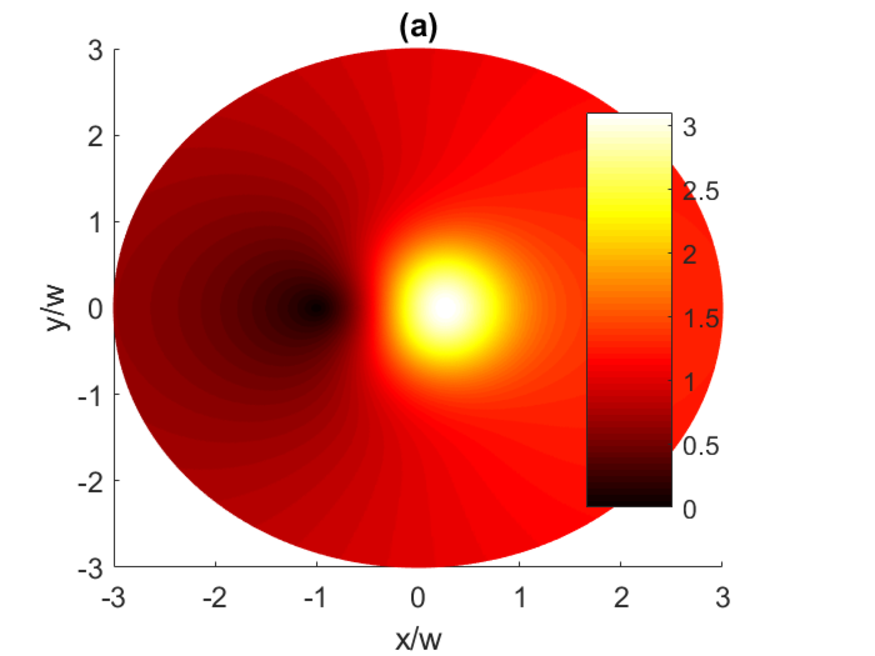}\includegraphics[width=0.3\textwidth]{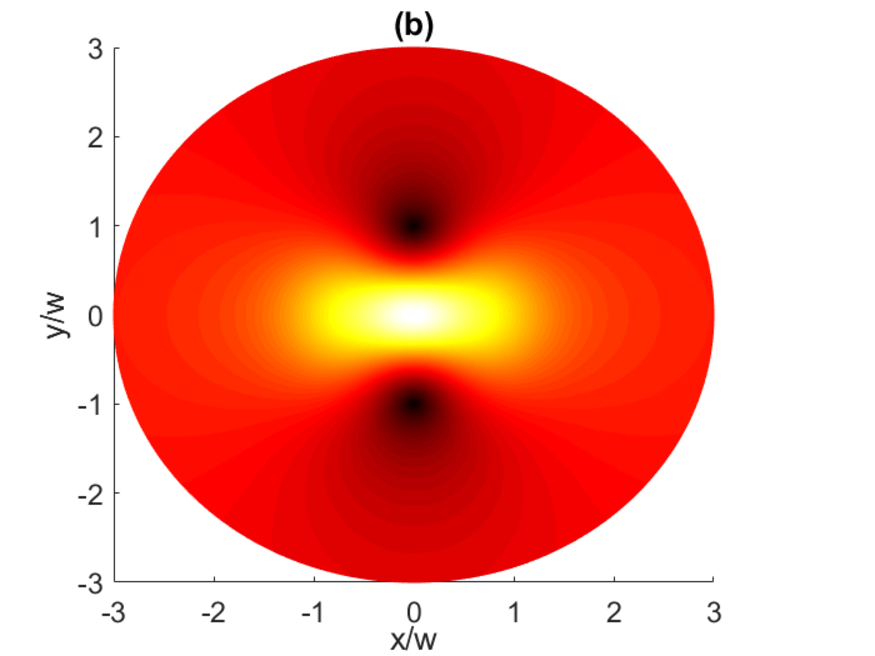}\includegraphics[width=0.3\textwidth]{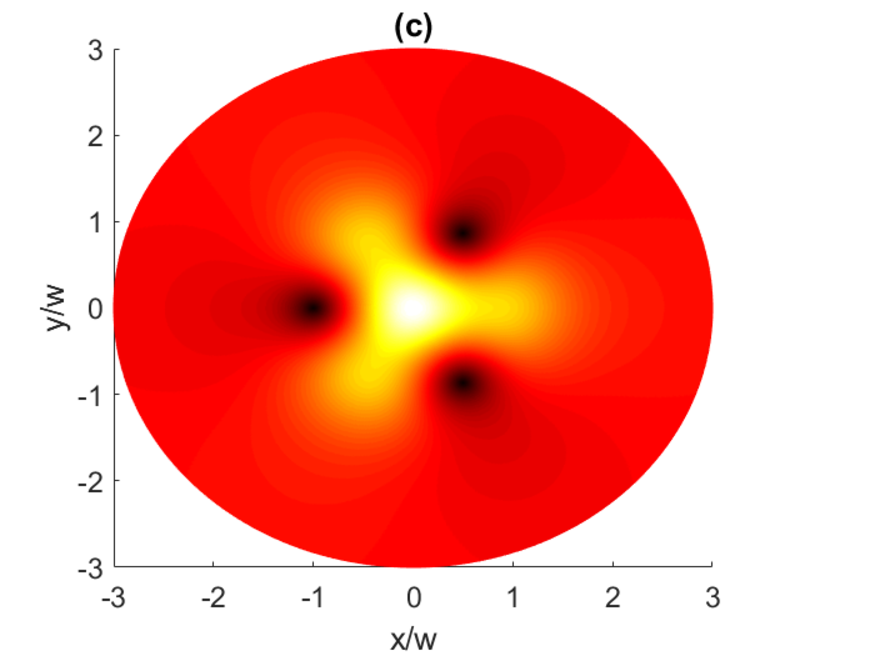}

\includegraphics[width=0.3\textwidth]{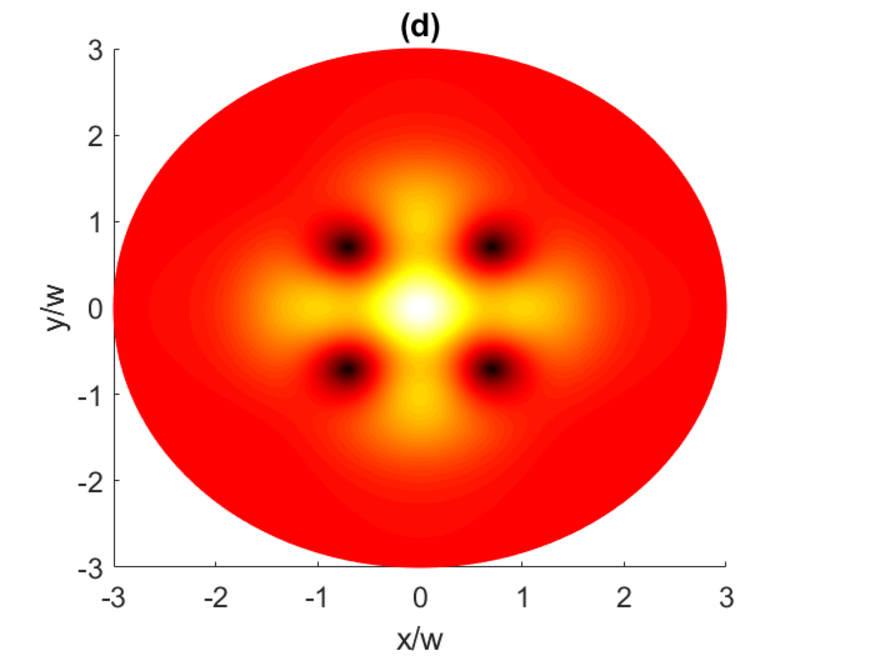}\includegraphics[width=0.3\textwidth]{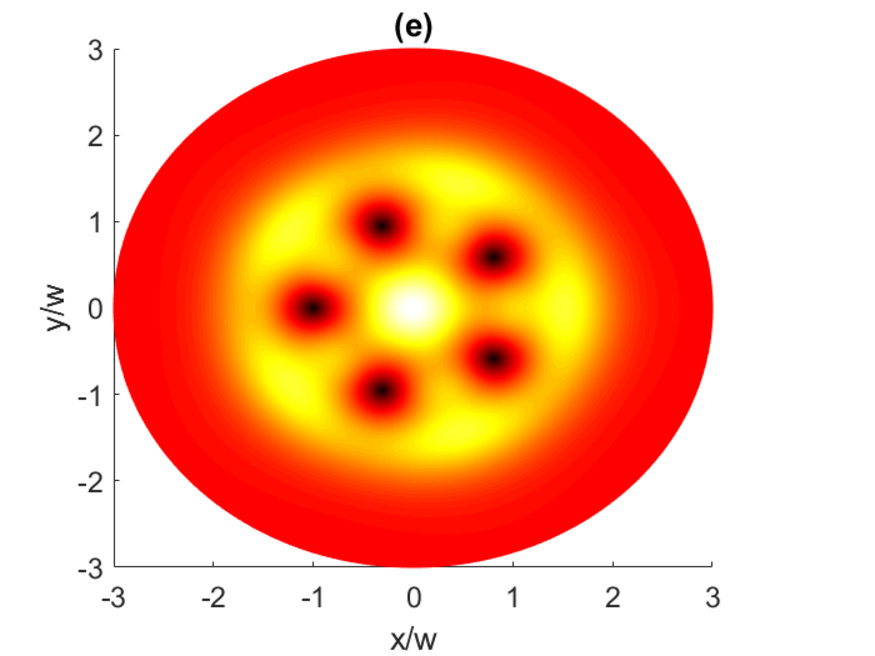}\includegraphics[width=0.3\textwidth]{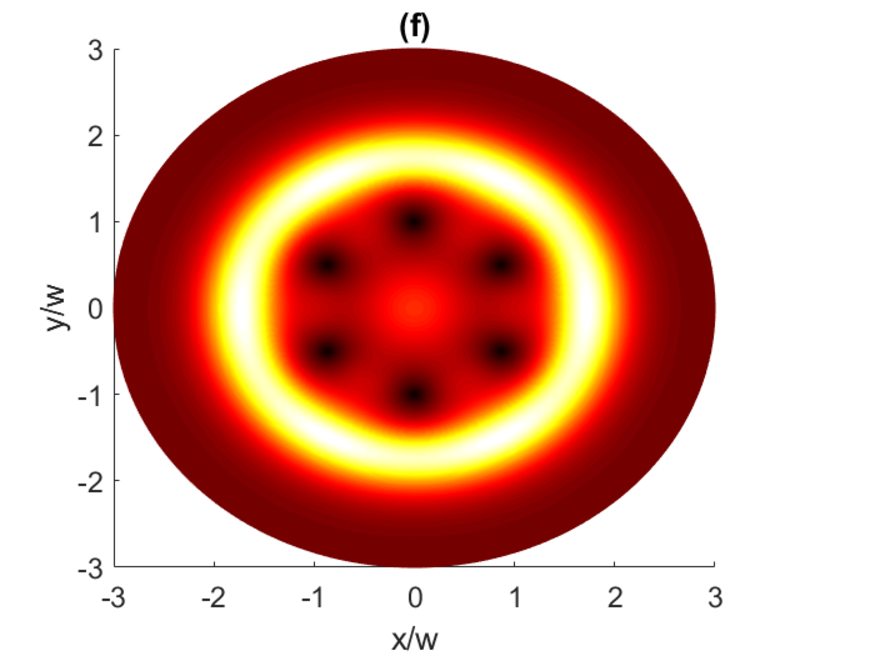}
\caption{Intensity distributions (in arbitrary units) of the subluminal superposition
beam $\psi$ featured in Eq.~(\ref{eq:25}) with different vorticities
$l_{1}=1-6$ (a)-(e). Here the parameters are $|\Omega_{c_{1}}|=|\Omega_{c_{2}}|=\Gamma$,
$\delta_{1f}=\Gamma$, $z=L_{\Gamma}$ and $\delta_{2f}=0$.}
\label{fig:4}
\end{figure}

\begin{figure}
\includegraphics[width=0.3\textwidth]{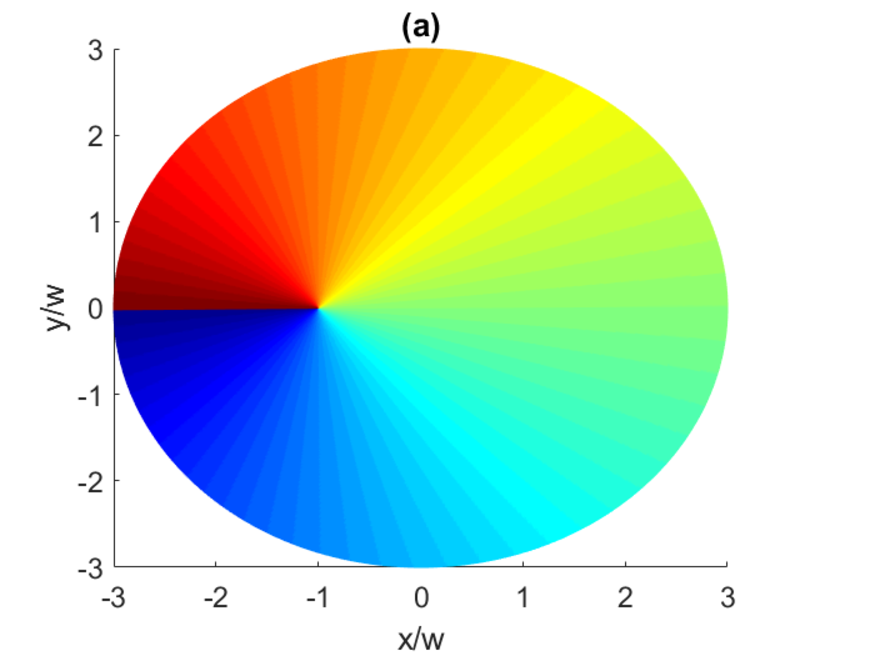}\includegraphics[width=0.3\textwidth]{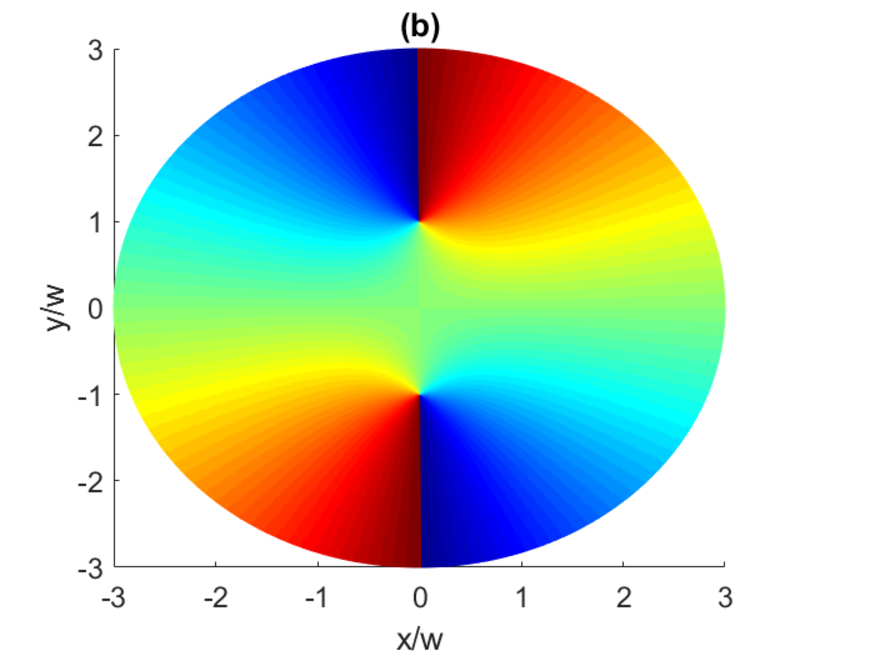}\includegraphics[width=0.3\textwidth]{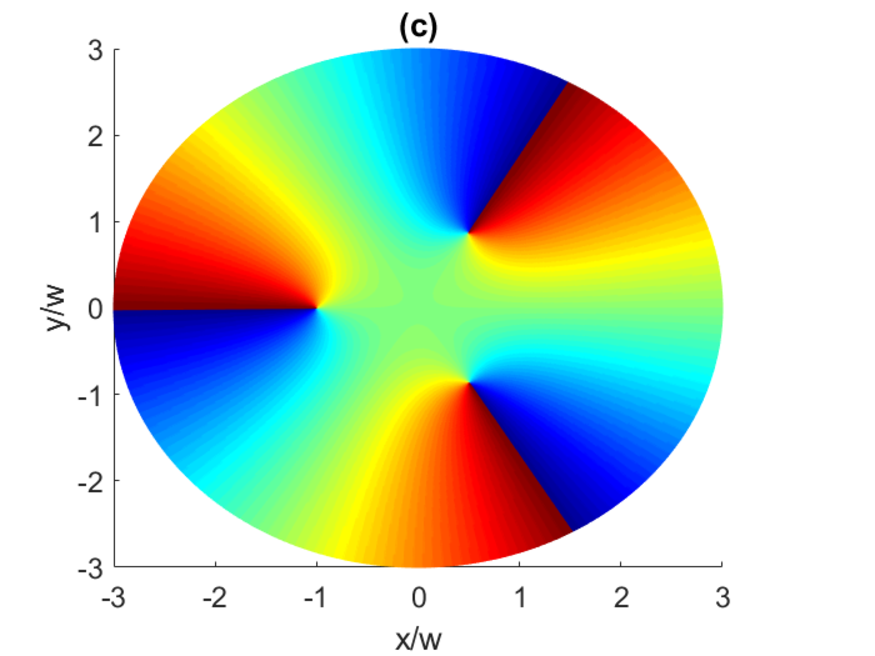}

\includegraphics[width=0.3\textwidth]{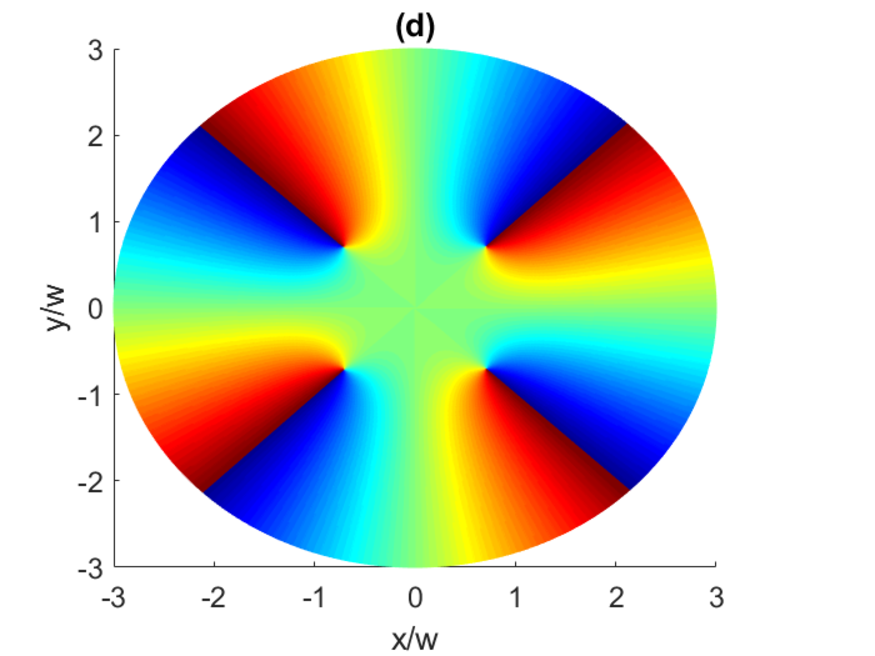}\includegraphics[width=0.3\textwidth]{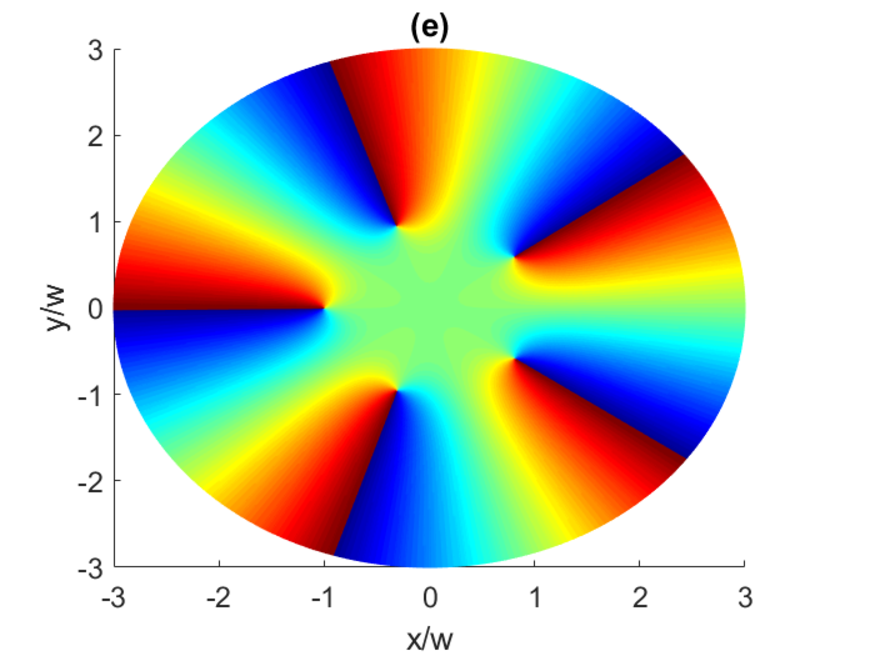}\includegraphics[width=0.3\textwidth]{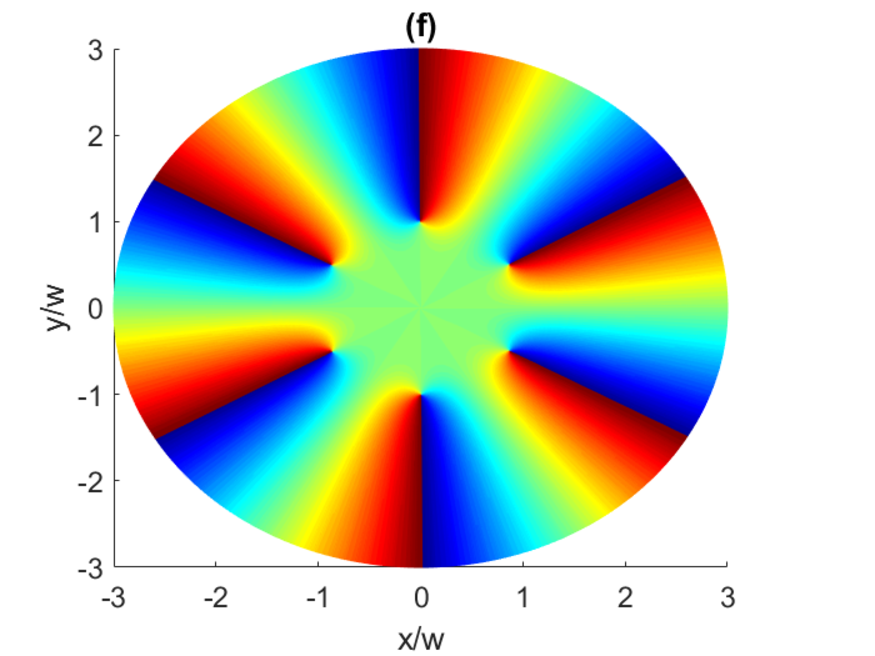}

\caption{The helical phase patterns of the subluminal superposition beam $\psi$
featured in Eq.~(\ref{eq:25}) with different vorticities $l_{1}=1-6$
(a)-(e). The parameters are the same as Fig.~\ref{fig:4}.}
\label{fig:5-1}
\end{figure}

\begin{figure}
\includegraphics[width=0.3\textwidth]{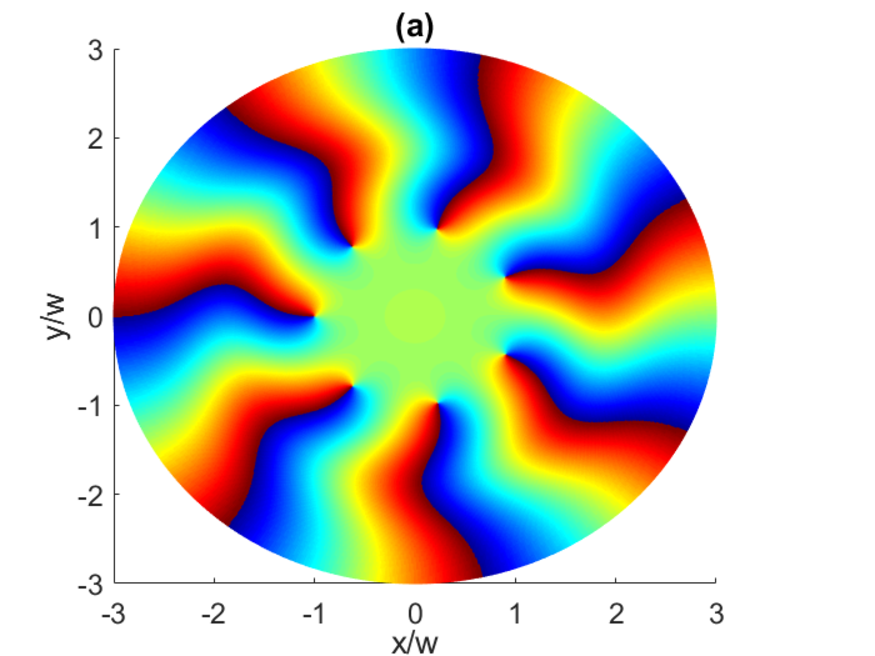}\includegraphics[width=0.3\textwidth]{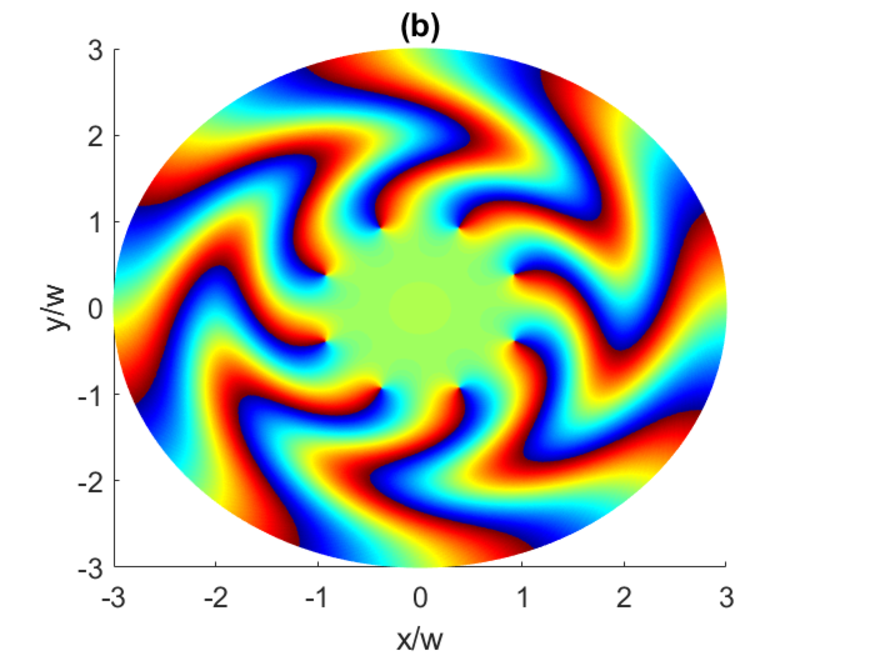}\includegraphics[width=0.3\textwidth]{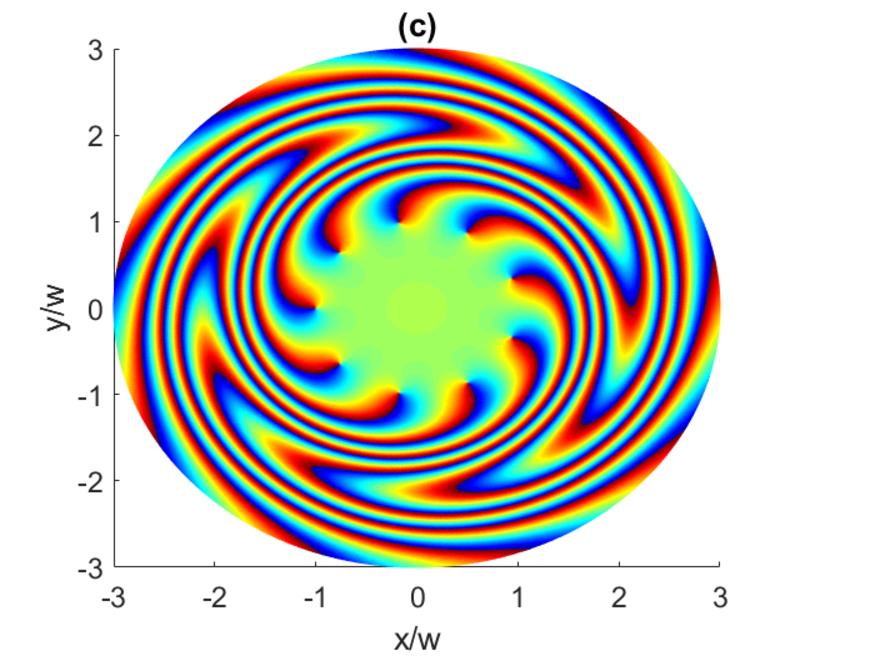}\includegraphics[width=0.3\textwidth]{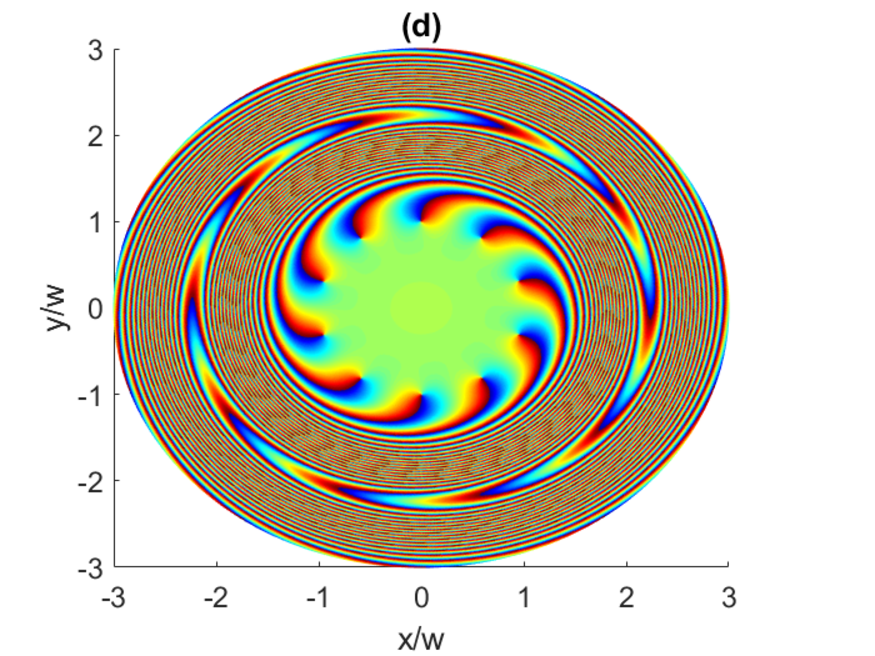}

\includegraphics[width=0.3\textwidth]{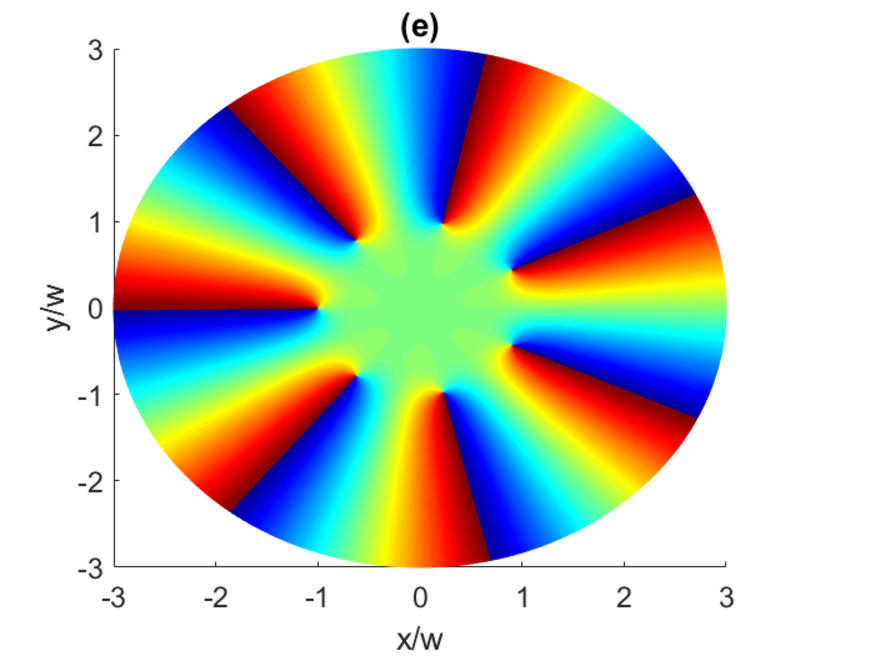}\includegraphics[width=0.3\textwidth]{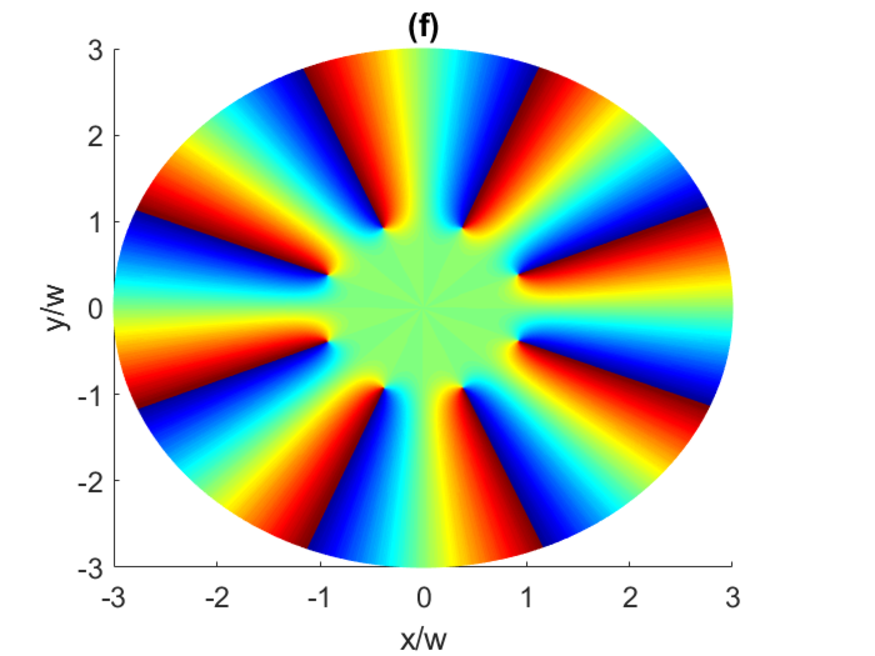}\includegraphics[width=0.3\textwidth]{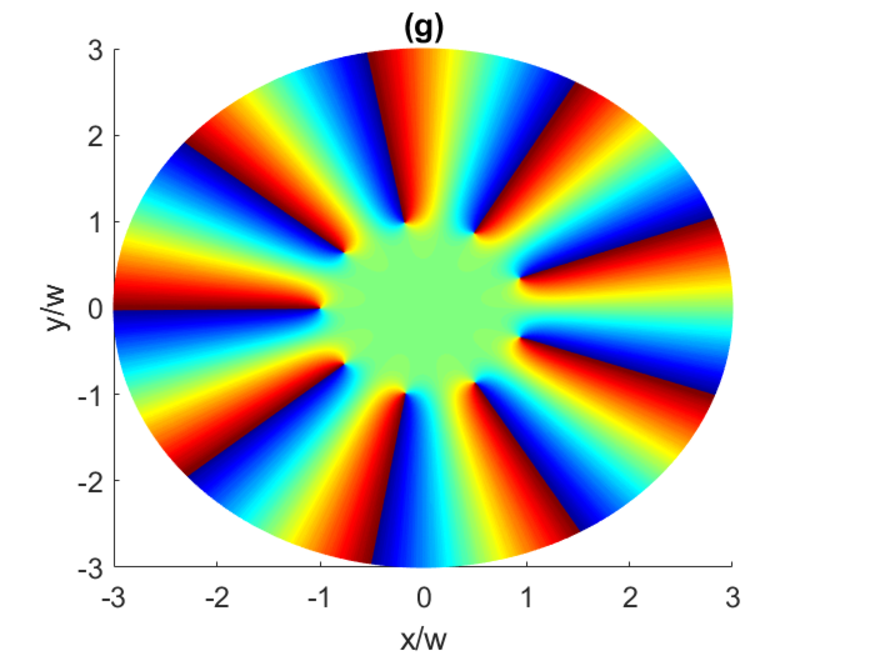}\includegraphics[width=0.3\textwidth]{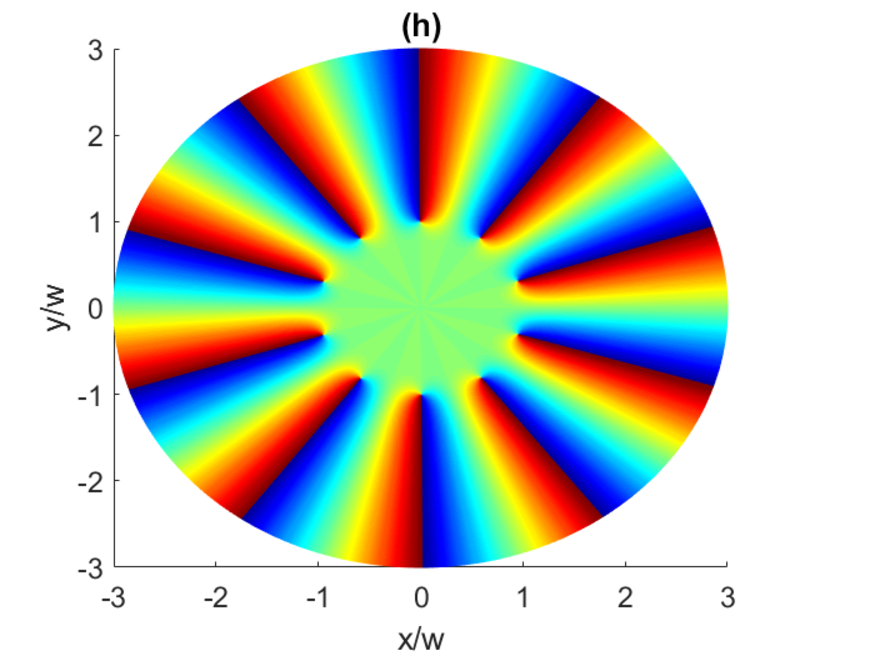}
\caption{The helical phase patterns of the superposition beam $\psi$ featured
in Eq.~(\ref{eq:25}) with different vorticities $l_{1}=7$ (a, e),
$l_{1}=8$ (b, f), $l_{1}=9$ (c, g) and $l_{1}=10$ (d, h). Here
$\delta_{2f}=4\Gamma$ (a,b,c,d), $\delta_{2f}=0$ (e,f,g,h) and the
other parameters are the same as in Fig.~(\ref{fig:2}).}
\label{fig6}
\end{figure}

\begin{figure}
\includegraphics[width=0.3\textwidth]{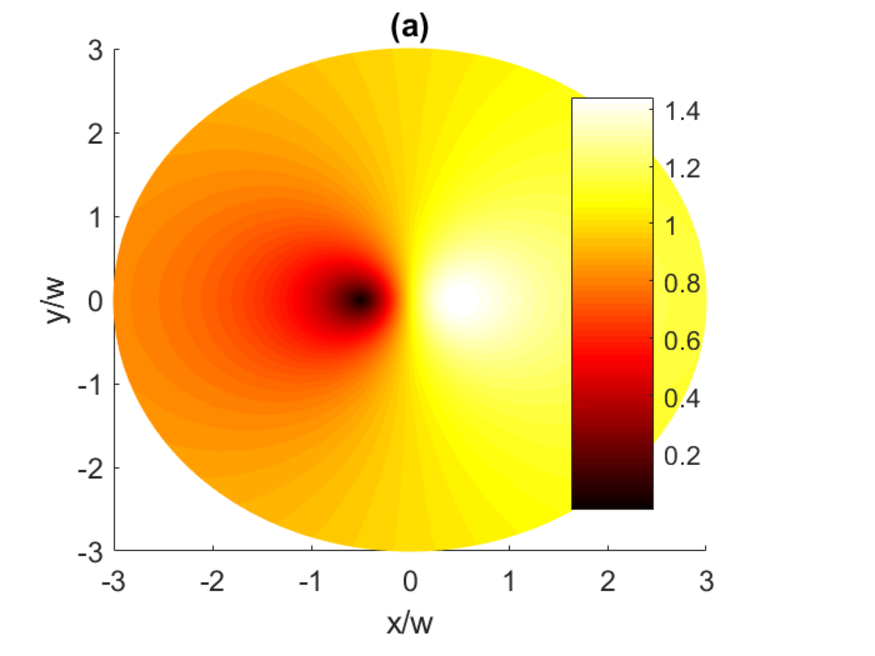}\includegraphics[width=0.3\textwidth]{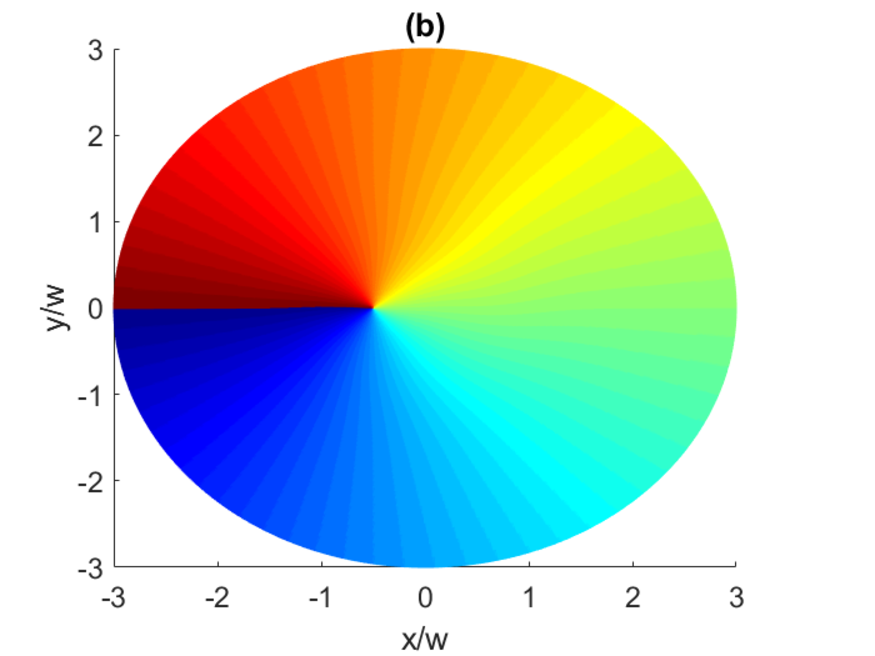}\includegraphics[width=0.3\textwidth]{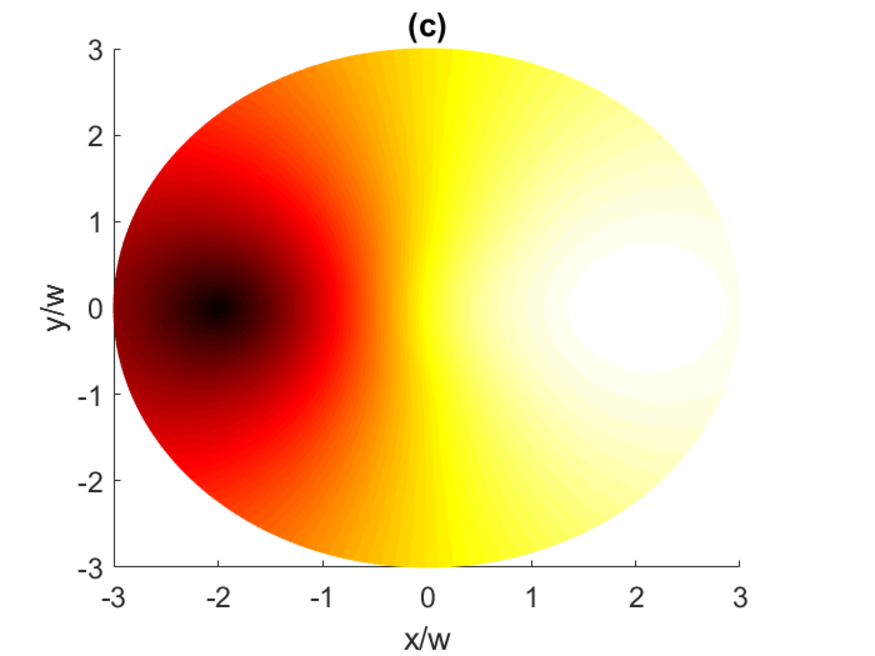}\includegraphics[width=0.3\textwidth]{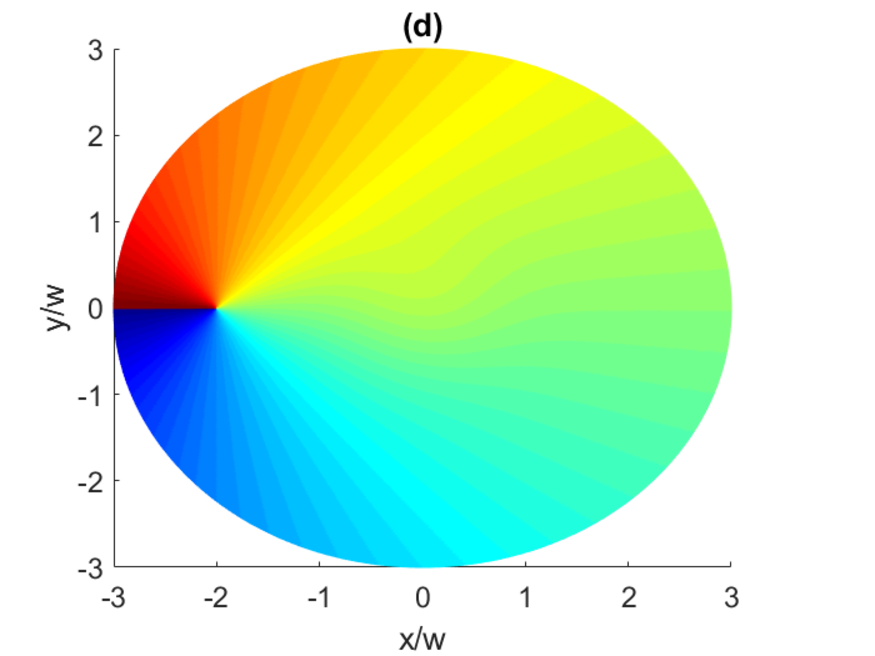}
\caption{Intensity distributions (a, c) in arbitrary units as well as the corresponding
helical phase patterns (b, d) of the superposition beam $\psi$ featured
in Eq.~(\ref{eq:25}). Here $|\Omega_{c_{1}}|=\Gamma$, $|\Omega_{c_{2}}|=0.5\Gamma$
(a, b) and $|\Omega_{c_{1}}|=0.5\Gamma$, $|\Omega_{c_{2}}|=\Gamma$
(c, d), $l_{1}=1$ and the other parameters are the same as in Fig.~(\ref{fig:2}).}
\label{fig7}
\end{figure}

In the following we consider a case where the first pump field $\Omega_{c_{1}}$
is a vortex $l_{1}\neq0$, while the second pump field is a non-vortex
Gaussian beam with $l_{2}=0$. We have made such an assumption to
avoid the zero denominator when $r\rightarrow0$ in Eq.~(\ref{eq:25})
if $l_{2}\neq0$. Numerical simulations presented in Figs.~(\ref{fig:2})-(\ref{fig7}) 
show the superposition beam given by Eq.~(\ref{eq:25}) in a 
transverse plane of the beam at $z=L_{\Gamma}$.

Figure~\ref{fig:2} (\ref{fig:4}) displays the numerical results
of the intensity distributions of the superposition beam $\psi$ when
the two-photon detuning is larger (smaller) than $\Gamma$ corresponding
to superluminal (subluminal) propagation of superposition pulse inside
the medium, and for different vorticities $l_{1}=1-6$. Figure~\ref{fig:3}
(\ref{fig:5-1}) shows the corresponding helical phase patterns. For
simulations we have selected $\delta_{2f}=4\Gamma$ and $\delta_{2f}=0$
corresponding to the superluminal and subluminal situations, respectively.

The resulting beam is seen to have a very particular shape.
The center of the superposition beam contains no vortex and is surrounded
by $l_{1}$ singly charged peripheral vortices of sign $l_{1}/|l_{1}|$.
The peripheral vortices are distributed at angles
\begin{equation}
\Theta_{p}=\frac{n\pi}{l_{1}},\label{eq:theta}
\end{equation}
with an approximate radial distance to the beam center
\begin{equation}
r_{p}\approx\left(|l_{1}|!\frac{|\Omega_{c_{2}}|}{|\Omega_{c_{1}}|}\right)^{\frac{1}{2|l_{1}|}},\label{eq:r}
\end{equation}
where $n=1...2|l_{1}|$ is an integer for each peripheral vortex \cite{Baumann2009}.
The off-axis vortices are placed at the same radial distance from
the core of the superposition beam.

Such images of the subluminal or superluminal vortices appear as two initial pump
beams with different azimuthal indices $l_{1}\neq0$ and $l_{2}=0$ are superimposed
leading to formation of the off-center vortices with
shifted axes. To elucidate this better, let us consider Fig.~\ref{fig:3}(f)
which is plotted for $l_{1}=6$. Note that we have considered a case
where the strength of both coupling beams are the same ($|\Omega_{c_{1}}|=|\Omega_{c_{2}}|=\Gamma$).
Region $A$ is dominated by the vortex beam with $\Omega_{c_{1}}$
and  $l_{1}=6$, while the inner region $B$ is dominated by the Gaussian
beam $\Omega_{c_{2}}$ with $l_{2}=0$. The peripheral vortices are
located precisely at the boundary between the two regions which is
a circle of the radius $r_{p}$.

Comparing of Figs.~\ref{fig:3} and \ref{fig:5-1} shows that the phase
structures of the superluminal superposition beam is bent with respect
to the subluminal one. Such a bending of the phase patterns becomes more
significant when the topological charge $l_{1}$ increases, as one can see comparing
Figs.~\ref{fig6} (a,b,c,d) with \ref{fig6} (e,f,g,h). In fact, the exponent of the factor $e^{i\frac{\Gamma}{\delta_{1f}^{2}}\frac{\mathcal{E}_{c}^{2}(r)}{(\delta_{2f}+i\Gamma)}\frac{z}{L_{\Gamma}}}$
in Eq.~(\ref{eq:25}) 
contains the term $\mathcal{E}_{c}^{2}(r)$ which is not uniform in the $(x,y)$ plane resulting
to bending of the phase patterns when $\delta_{2f}$ is nonzero.

Figure~\ref{fig7} illustrates the effect of the strength of pump
beams $|\Omega_{c_{1}}|$ and $|\Omega_{c_{2}}|$ on intensity distributions
and the corresponding helical phase patterns. We plot only the case
of the superluminality ($\delta_{2f}=4\Gamma$), as the results are very
similar to the subluminal case. It is apparent from Fig.~\ref{fig7}
(a, b) that the peripheral vortex shifts toward the center of the
beam when $|\Omega_{c_{1}}|>|\Omega_{c_{2}}|$ while it moves away
from the core when $|\Omega_{c_{1}}|<|\Omega_{c_{2}}|$ (see Fig.~\ref{fig7}
(c, d)). As can be seen from Eq.~(\ref{eq:r}), when $|\Omega_{c_{1}}|>|\Omega_{c_{2}}|$
( $|\Omega_{c_{1}}|<|\Omega_{c_{2}}|$), the radius $r_{p}$ reduces (increases) and the position of
the peripheral vortex moves radially in (out).

\subsection{Exchange of optical vortices }

\begin{figure}
\includegraphics[width=0.3\textwidth]{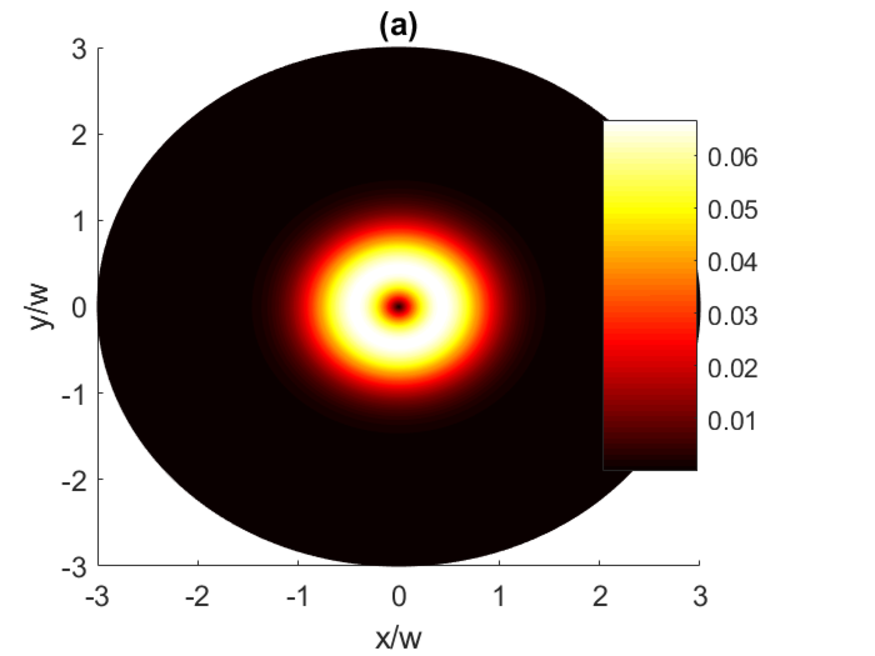}\includegraphics[width=0.3\textwidth]{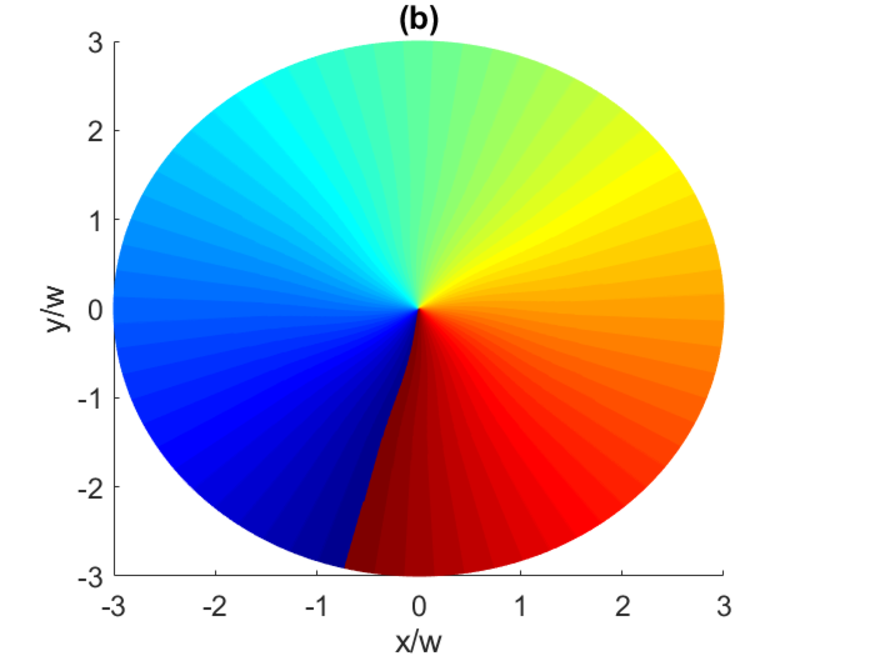}\includegraphics[width=0.3\textwidth]{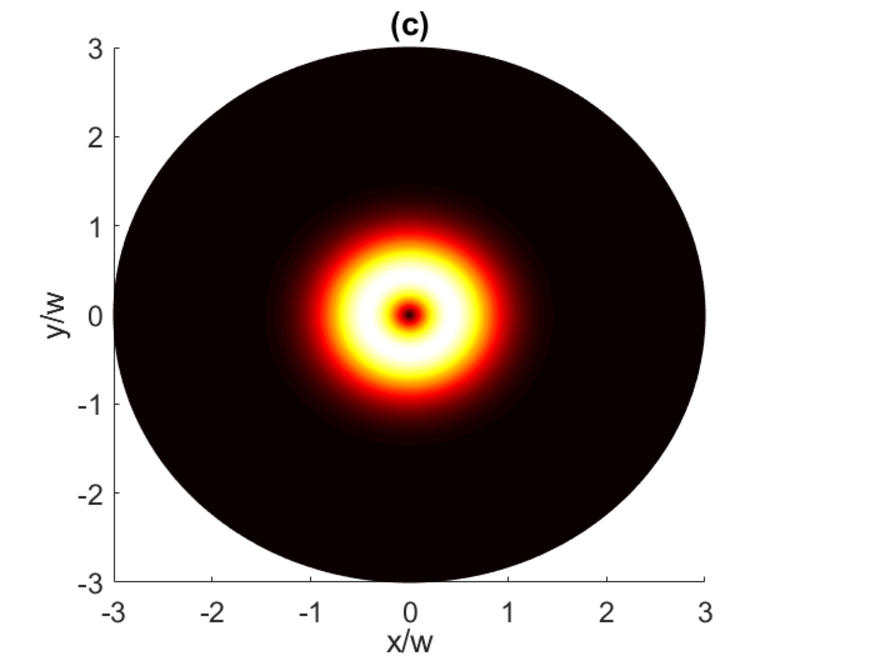}\includegraphics[width=0.3\textwidth]{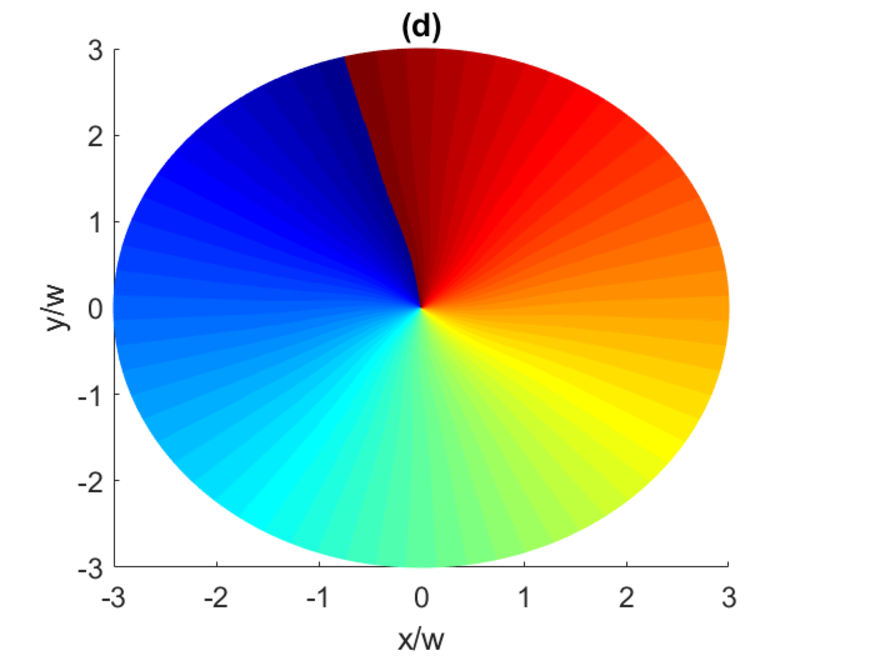}

\includegraphics[width=0.3\textwidth]{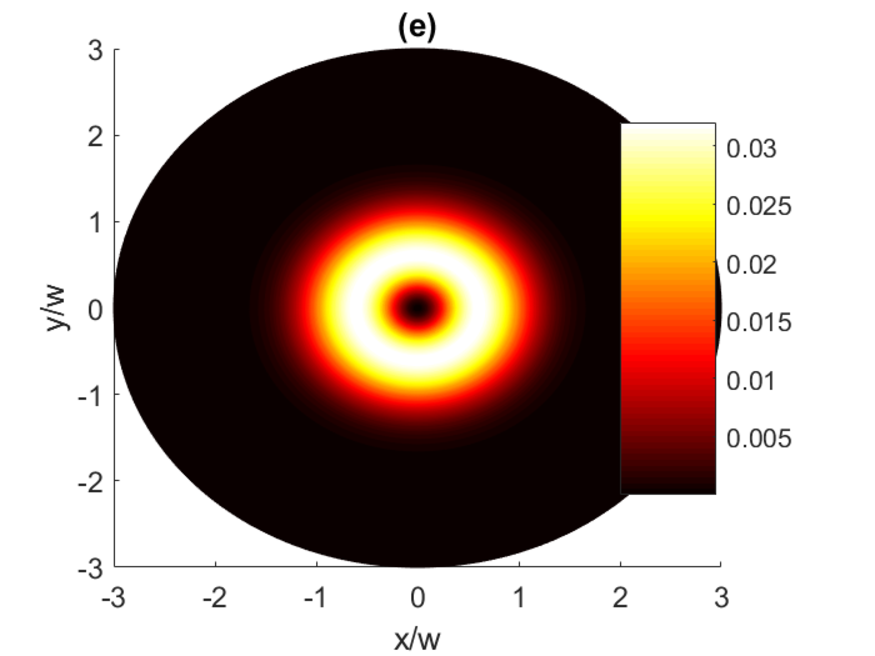}\includegraphics[width=0.3\textwidth]{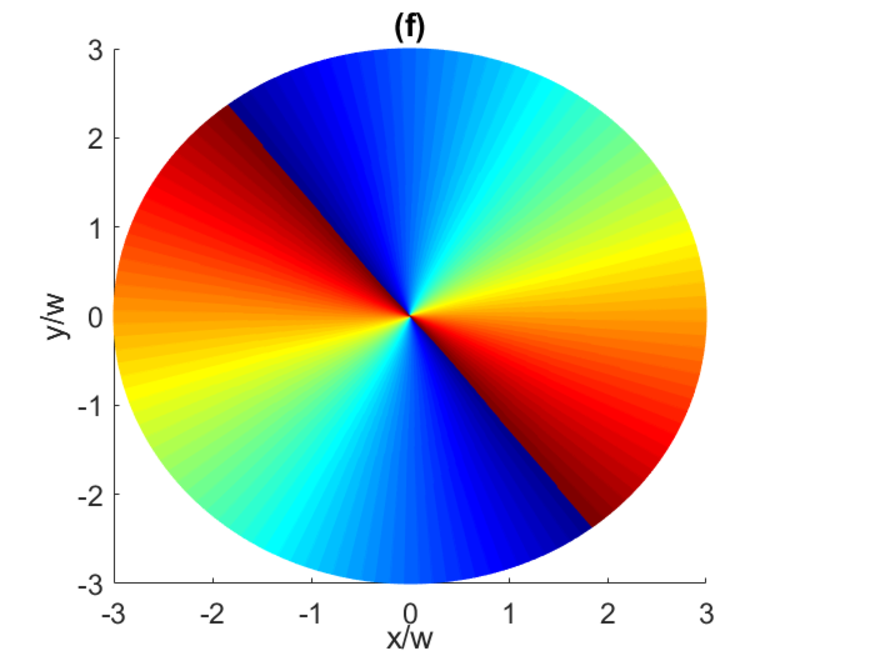}\includegraphics[width=0.3\textwidth]{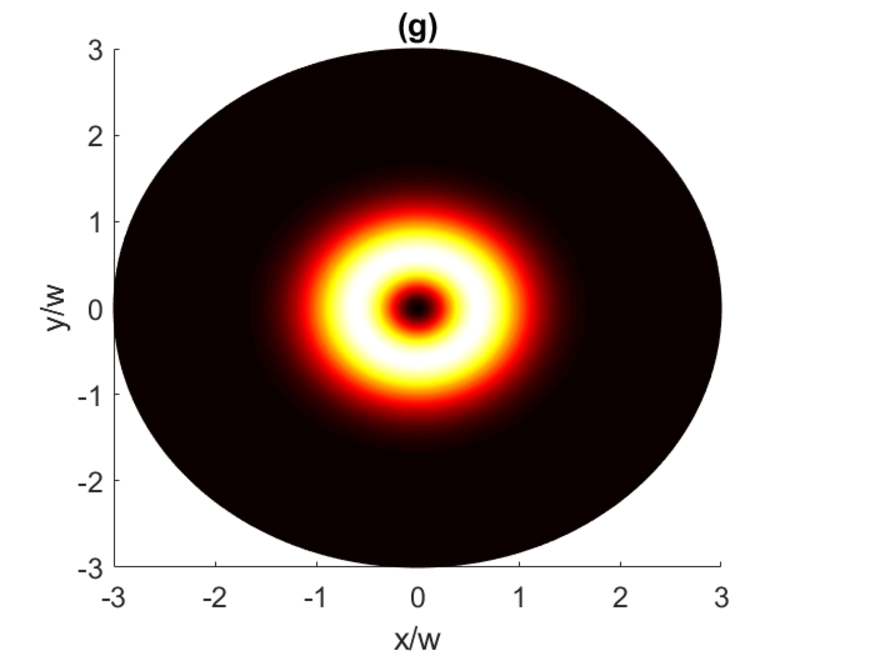}\includegraphics[width=0.3\textwidth]{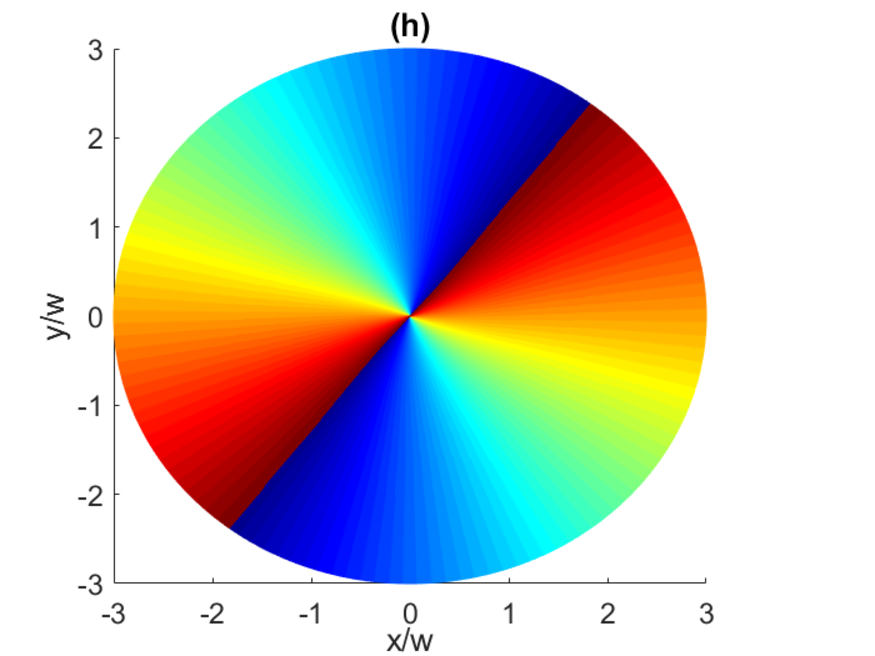}

\includegraphics[width=0.3\textwidth]{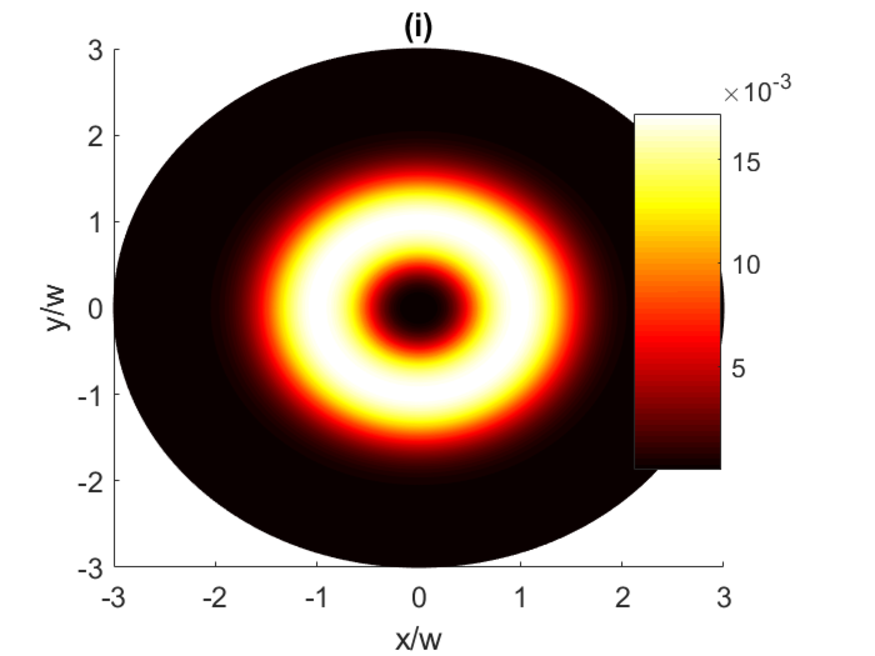}\includegraphics[width=0.3\textwidth]{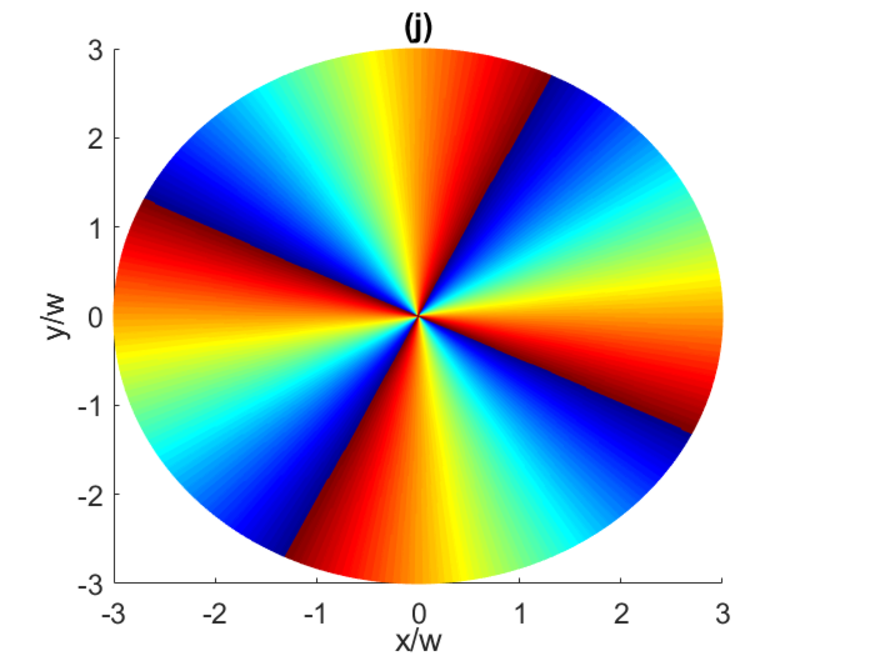}\includegraphics[width=0.3\textwidth]{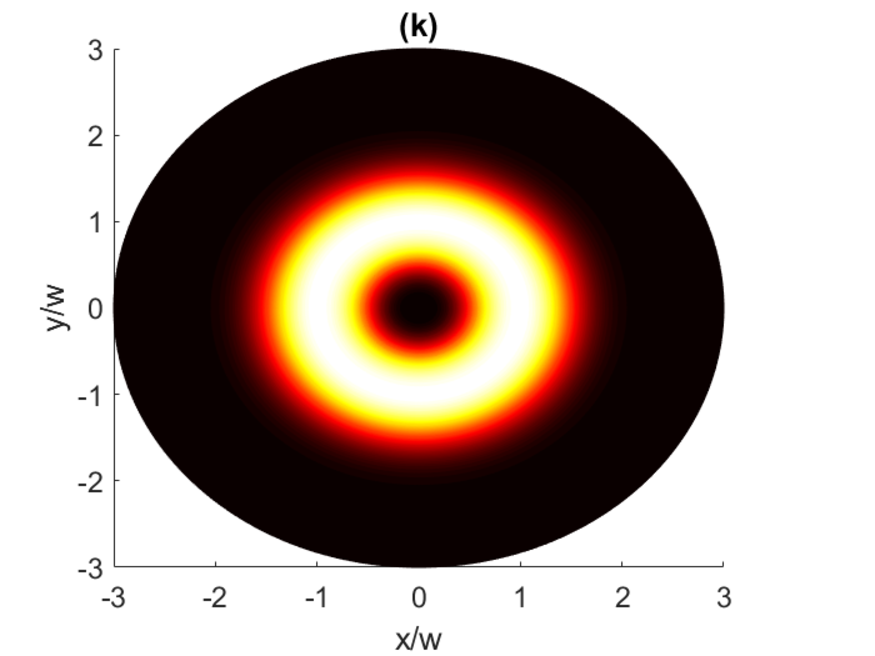}\includegraphics[width=0.3\textwidth]{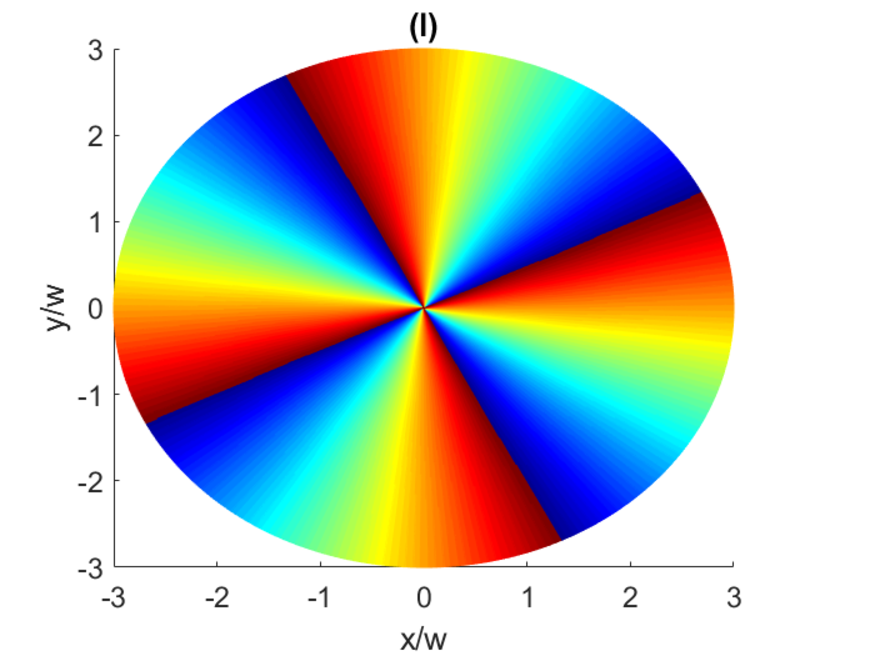}
\caption{Intensity distributions (a, c, e, g,i,k) in arbitrary units as well
as the corresponding helical phase patterns (b, d, f, h,j,l) of the
generated second probe vortex beam $\mathcal{P}_{2}$ featured in
Eq.~(\ref{eq:45-1}) when $l_{1}=1$, $l_{2}=0$ (a, b), $l_{1}=0$,
$l_{2}=1$ (c, d), $l_{1}=2$, $l_{2}=0$ (e, f), $l_{1}=0$, $l_{2}=1$
(g, h) and $l_{1}=4$, $l_{2}=0$ (i, j), $l_{1}=0$, $l_{2}=4$ (k,
l). Here $\delta_{2f}=4\Gamma$ and the other parameters are the same
as Fig.~(\ref{fig:2}).}
\label{fig8}
\end{figure}

We will now assume that only one probe field $\mathcal{P}_{1}$ is
initially incident on the atomic cloud $z=0$ ($\mathcal{P}_{1}(0)=\mathcal{P}$).
The amplitude of the second probe field is zero at the beginning ($\mathcal{P}_{2}(0)=0$).
In this case, Eqs.~(\ref{eq:19}) and (\ref{eq:20}) reduce to
\begin{align}
\psi(0) & =\frac{1}{\mathcal{E}_{c}(r)}\left(\mathcal{E}_{c_{1}}^{*}(r)e^{-il_{1}\varphi}\mathcal{P}\right),\label{eq:42-2}\\
\xi(0) & =\frac{1}{\mathcal{E}_{c}(r)}\left(\mathcal{E}_{c_{2}}(r)e^{il_{2}\varphi}\mathcal{P}\right).\label{eq:43-1}
\end{align}
The electric fields of the probe beams inside the atomic cloud can
be obtained from the fields $\psi$ and $\xi$ as
\begin{align}
\mathcal{P}_{1}(z) & =\frac{1}{\mathcal{E}_{c}(r)}\left(\mathcal{E}_{c_{1}}(r)e^{il_{1}\varphi}\psi(z)+\mathcal{E}_{c_{2}}^{*}(r)e^{-il_{2}\varphi}\xi(z)\right)=\left(1+\frac{\mathcal{E}_{c_{1}}^{2}(r)}{\mathcal{E}_{c}^{2}(r)}(e^{i\kappa z}-1)\right)\mathcal{P},\label{eq:44-1}\\
\mathcal{P}_{2}(z) & =\frac{1}{\mathcal{E}_{c}(r)}\left(\mathcal{E}_{c_{2}}(r)e^{il_{2}\varphi}\psi(z)-\mathcal{E}_{c_{1}}^{*}(r)e^{-il_{1}\varphi}\xi(z)\right)=\frac{\mathcal{E}_{c_{2}}(r)\mathcal{E}_{c_{1}}^{*}(r)}{\mathcal{E}_{c}^{2}(r)}e^{i(l_{2}-l_{1})}(e^{i\kappa z}-1)\mathcal{P}.\label{eq:45-1}
\end{align}
where $\kappa$ is given by Eq.~(\ref{eq:24}). The intensity
distributions and the corresponding helical phase pattern of the generated
second probe vortex beam are shown in Fig.~\ref{fig8} for 
$\delta_{2f}=4\Gamma$ and different
topological charge numbers. A doughnut
intensity profile is observed with a dark hollow in the center. The phase
jumps from $0$ to $n\pi$ around the singularity point. As Eq.~(\ref{eq:45-1}) shows, the generated field contains a phase factor of $e^{i(l_{2}-l_{1})}$. If the first
pump field is a vortex but the second one is a non-vortex beam, the generated probe field acquires a vortex of charge $-l_{1}$. On the other hand, if only the second pump beam is a vortex with the charge $l_{2}$, the generated probe beam has a vorticity $l_{2}$.

\section{The double Raman doublet scheme}

\begin{figure}
\includegraphics[width=0.4\textwidth]{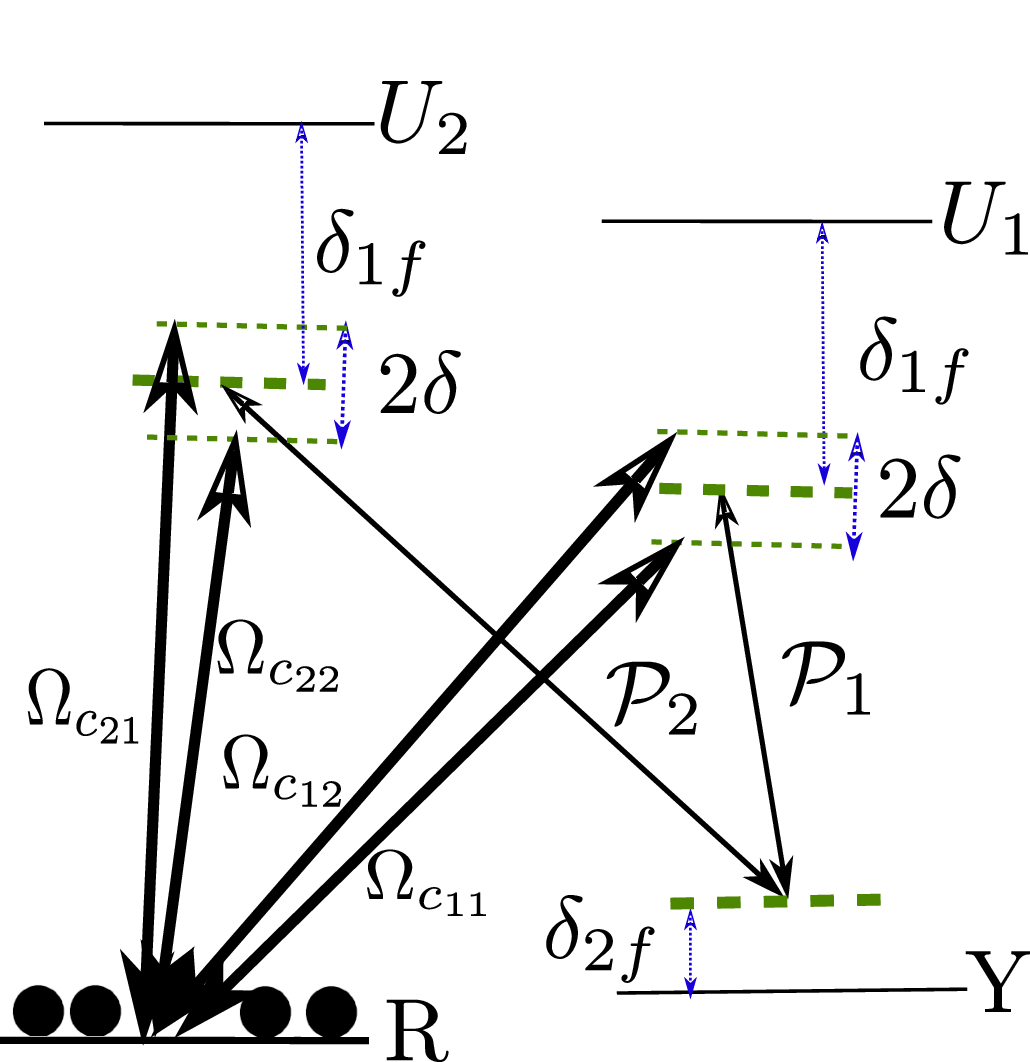}
\caption{Schematic diagram of the double-Raman doublet scheme.}
\label{fig:DRD}
\end{figure}

In this section we present a more favorable scenario for the generation
of off-axis vortices. We consider a situation where
four strong pump beams act on the atomic ensemble (Fig.~(\ref{fig:DRD})).
This situation corresponds to a Raman doublet for each of the probe
beams.

We assume four-photon resonances $\omega_{c_{11}}-\omega_{p_{1}}=\omega_{c_{21}}-\omega_{p_{2}}$,
$\omega_{c_{12}}-\omega_{p_{1}}=\omega_{c_{22}}-\omega_{p_{2}}$,
where $\omega_{c_{11}},$$\omega_{c_{12}}$, $\omega_{c_{21}}$ and
$\omega_{c_{22}}$ are frequencies of the pump beams. To describe
the propagation of the probe beams in the medium, we separate the
atomic amplitudes into two parts oscillating with different frequencies:
$\textbf{\ensuremath{\Psi_{Y}}}=\textbf{\ensuremath{\Psi_{Y_{1}}+\textbf{\ensuremath{\Psi_{Y_{2}}}}}}$,
$\textbf{\ensuremath{\Psi_{U_{1}}}}=\textbf{\ensuremath{\Psi_{U_{11}}+\textbf{\ensuremath{\Psi_{U_{12}}}}}}$
and $\textbf{\ensuremath{\Psi_{U_{2}}}}=\textbf{\ensuremath{\Psi_{U_{21}}+\textbf{\ensuremath{\Psi_{U_{22}}}}}}$.
Recalling the slowly changing amplitudes and neglecting the terms
oscillating with a large frequency $2\delta=\omega_{c_{12}}-\omega_{c_{11}}=\omega_{c_{22}}-\omega_{c_{21}}$,
the equations for the matter fields read
\begin{align}
c\partial_{z}\mathcal{P}_{1}= & i\alpha\Phi_{Y_{1}}^{*}\frac{\mathcal{E}_{c_{11}}(r)}{\delta_{1f}}e^{il_{11}\varphi}\Phi_{R}+i\alpha\Phi_{Y_{2}}^{*}\frac{\mathcal{E}_{c_{12}}(r)}{\delta_{1f}}e^{il_{12}\varphi}\Phi_{R},\label{eq:28}\\
c\partial_{z}\mathcal{P}_{2}= & i\alpha\Phi_{Y_{1}}^{*}\frac{\mathcal{E}_{c_{21}}(r)}{\delta_{1f}}e^{il_{21}\varphi}\Phi_{R}+i\alpha\Phi_{Y_{2}}^{*}\frac{\mathcal{E}_{c_{22}}(r)}{\delta_{1f}}e^{il_{22}\varphi}\Phi_{R},\label{eq:29}\\
\Phi_{Y_{1}}= & \frac{\alpha\Phi_{R}}{(\delta_{2f}+\delta-i\Gamma)\delta_{1f}}\left(\mathcal{E}_{c_{11}}(r)e^{il_{11}\varphi}\mathcal{P}_{1}^{*}+\mathcal{E}_{c_{21}}(r)e^{il_{21}\varphi}\mathcal{P}_{2}^{*}\right),\label{eq:30}\\
\Phi_{Y_{2}}= & \frac{\alpha\Phi_{R}}{(\delta_{2f}-\delta-i\Gamma)\delta_{1f}}\left(\mathcal{E}_{c_{12}}(r)e^{il_{12}\varphi}\mathcal{P}_{1}^{*}+\mathcal{E}_{c_{22}}(r)e^{il_{22}\varphi}\mathcal{P}_{2}^{*}\right),\label{eq:31}
\end{align}
where $\delta_{1f}=\omega_{U_{1}}-\omega_{R}-\frac{1}{2}(\omega_{c_{11}}+\omega_{c_{12}})=\omega_{U_{2}}-\omega_{R}-\frac{1}{2}(\omega_{c_{21}}+\omega_{c_{22}})$
is an average one-photon detuning and $\delta_{2f}=\omega_{p_{1}}+\omega_{Y}-\omega_{R}-\frac{1}{2}(\omega_{c_{11}}+\omega_{c_{12}})=\omega_{p_{2}}+\omega_{Y}-\omega_{R}-\frac{1}{2}(\omega_{c_{21}}+\omega_{c_{22}})$
denotes an average two-photon detuning. 
 Equations (\ref{eq:28})-(\ref{eq:31}) give the following equations for the propagation of the probe fields 
\begin{align}
\partial_{z}P_{1}-i\beta\left(\frac{|\mathcal{E}_{c_{11}}(r)|^{2}\mathcal{P}_{1}+\mathcal{E}_{c_{11}}(r)\mathcal{E}_{c_{21}}^{*}(r)e^{i(l_{11}-l_{21})\varphi}\mathcal{P}_{2}}{\delta_{2f}+\delta+i\Gamma}+\frac{|\mathcal{E}_{c_{12}}(r)|^{2}\mathcal{P}_{1}+\mathcal{E}_{c_{12}}(r)\mathcal{E}_{c_{22}}^{*}(r)e^{i(l_{12}-l_{22})\varphi}\mathcal{P}_{2}}{\delta_{2f}-\delta+i\Gamma}\right) & =0,\label{eq:32}\\
\partial_{z}P_{2}-i\beta\left(\frac{\mathcal{E}_{c_{21}}(r)\mathcal{E}_{c_{11}}^{*}(r)e^{i(l_{21}-l_{11})\varphi}\mathcal{P}_{1}+|\mathcal{E}_{c_{21}}(r)|^{2}\mathcal{P}_{2}}{\delta_{2f}+\delta+i\Gamma}+\frac{\mathcal{E}_{c_{22}}(r)\mathcal{E}_{c_{12}}^{*}(r)e^{i(l_{22}-l_{12})\varphi}\mathcal{P}_{1}+|\mathcal{E}_{c_{22}}(r)|^{2}\mathcal{P}_{2}}{\delta_{2f}-\delta+i\Gamma}\right) & =0.\label{eq:33}
\end{align}

We consider a particular situation in which
\begin{equation}
\frac{\mathcal{E}_{c_{12}}(r)}{\mathcal{E}_{c_{11}}(r)}e^{i(l_{12}-l_{11})\varphi}=\frac{\mathcal{E}_{c_{22}}(r)}{\mathcal{E}_{c_{21}}(r)}e^{i(l_{22}-l_{21})\varphi}.\label{eq:34}
\end{equation}
Defining the generalized quantities
\begin{align}
\mathcal{E}_{c_{1}}(r) & =\sqrt{|\mathcal{E}_{c_{11}}(r)|^{2}+|\mathcal{E}_{c_{21}}(r)|^{2}},\label{eq:38}\\
\mathcal{E}_{c_{2}}(r) & =\sqrt{|\mathcal{E}_{c_{12}}(r)|^{2}+|\mathcal{E}_{c_{22}}(r)|^{2}},\label{eq:39}
\end{align}
and introducing new fields representing superpositions of the original
probe fields
\begin{equation}
\psi=\frac{1}{\mathcal{E}_{c_{1}}(r)}\left(\mathcal{E}_{c_{11}}^{*}(r)e^{-il_{11}\varphi}\mathcal{P}_{1}+\mathcal{E}_{c_{21}}^{*}(r)e^{-il_{21}\varphi}\mathcal{P}_{2}\right),\label{eq:36}
\end{equation}
\begin{equation}
\xi=\frac{1}{\mathcal{E}_{c_{1}}(r)}\left(\mathcal{E}_{c_{21}}(r)e^{il_{21}\varphi}\mathcal{P}_{1}-\mathcal{E}_{c_{11}}(r)e^{il_{11}\varphi}\mathcal{P}_{2}\right),\label{eq:37}
\end{equation}

reduce Eqs.~(\ref{eq:32}) and (\ref{eq:33}) to Eqs.~(\ref{eq:22})
and (\ref{eq:23}) with
\begin{equation}
\kappa=\beta\left(\frac{\mathcal{E}_{c_{1}}^{2}(r)}{(\delta_{2f}+\delta+i\Gamma)}+\frac{\mathcal{E}_{c_{2}}^{2}(r)}{(\delta_{2f}-\delta+i\Gamma)}\right).\label{eq:40}
\end{equation}
Again, the field $\xi$ does not interact and propagates as in free
space, while the new field $\psi$ interacts with the atoms. Assuming
\begin{equation}
\mathcal{E}_{c_{1}}(r)=\mathcal{E}_{c_{2}}(r)=\mathcal{E}_{c}(r),\label{eq:44a}
\end{equation}
 and $\delta_{2f}=0$, one finds
\begin{equation}
\nu_{g}=\frac{c}{1+\frac{2\alpha^{2}n\mathcal{E}_{c}^{2}(r)}{\delta_{1f}^{2}}\frac{\Gamma^{2}-\delta^{2}}{(\delta^{2}+\Gamma^{2})^{2}}}.\label{eq:41}
\end{equation}
Thus for $\delta>\Gamma$ ($\delta<\Gamma$) the group velocity is
larger (smaller) than $c$ providing superluminal (superluminal) propagation.  
 In addition, according to the Eq.~(\ref{eq:40}) , the generated fast light experiences again amplification. We
see that, in contrast to the double-Raman scheme, we have sup- or
superluminal propagation even for zero two-photon detuning $\delta_{2f}=0$.
In order to have off-axis optical vortices satisfying Eqs.~(\ref{eq:34}),
(\ref{eq:38}), (\ref{eq:39}), (\ref{eq:44a}) and to avoid zero
denominator at the core in $\psi(z)=\psi(0)e^{i\kappa z}$, we can
consider $l_{11}=-l_{22}=l\neq0$ (i.e. $\Omega_{c_{11}}$ and $\Omega_{c_{22}}$
are vortices) but $l_{21}=l_{12}=0$ (i.e. $\Omega_{c_{11}}$ and
$\Omega_{c_{22}}$ are non-vortex Gaussian beams).

Let us assume that only one probe field $\mathcal{P}_{1}$ is incident
on the atomic cloud ($\mathcal{P}_{1}(0)=\mathcal{P}$). The amplitude
of the second probe field at the beginning of the atomic cloud $z=0$
is zero ($\mathcal{P}_{2}(0)=0$). In this case, Eqs.~(\ref{eq:36})
and (\ref{eq:37}) reduce to
\begin{align}
\psi(0) & =\frac{\mathcal{E}_{c_{11}}^{*}(r)e^{-il\varphi}}{\mathcal{E}_{c_{1}}(r)}\mathcal{P},\label{eq:42}\\
\xi(0) & =\frac{\mathcal{E}_{c_{21}}(r)}{\mathcal{E}_{c_{1}}(r)}\mathcal{P}.\label{eq:43}
\end{align}
The electric fields of the probe beams inside the atomic cloud can
be obtained from the fields $\psi$ and $\xi$ as
\begin{align}
\mathcal{P}_{1}(z) & =\left(1+\frac{\mathcal{E}_{c_{11}}^{2}(r)}{\mathcal{E}_{c}^{2}(r)}(e^{i\kappa z}-1)\right)\mathcal{P},\label{eq:44}\\
\mathcal{P}_{2}(z) & =\frac{\mathcal{E}_{c_{21}}(r)\mathcal{E}_{c_{11}}^{*}(r)}{\mathcal{E}_{c}^{2}(r)}e^{-il\varphi}(e^{i\kappa z}-1)\mathcal{P},\label{eq:45}
\end{align}
with $\kappa$  featured in Eq.~(\ref{eq:40}). Exchange of optical
vortices with opposite vorticity is now possible between the pump field $\Omega_{c_{11}}$
and the generated probe field $\mathcal{P}_{2}$ even for zero two-photon
detuning $\delta_{2f}=0$.

\section{Summary\label{sec:concl}}

We have investigated the formation of  off-axis vortices with shifted
axes in a double-Raman gain medium interacting with two weak probe
fields as well as two stronger pump lasers which can contain an optical
vortex. In such a medium only a particular superposition of the probe
fields is coupled with the atoms, while an orthogonal combination of the probe
fields does not interact with the atoms and propagates as in the free space.
Assuming that one of the pump fields is a vortex, depending on the
two-photon detuning, the superposition off-axis vortex beam can propagate
either with the slow or the fast group velocity. One can control the position
of the peripheral vortices by the vorticity and intensity of the pump fields.
The model for creation of the off-center fast and slow light
vortices can also be generalized to a more complicated double Raman doublet
with four pump fields. A possible experimental realization of the proposed scheme for off-axis
optical vortices can be implemented for an atomic cesium vapor cell
at the room temperature. All cesium atoms are to be prepared in the
ground-state hyperfine magnetic sublevel $6S_{1/2},|F=4,m=-4\rangle$
serving as the level $R$ in our scheme. The magnetic sublevel $6S_{1/2},|F=4,m=-2\rangle$
corresponds to the level $Y$. Also, the levels $5P_{3/2},|F=4,m=-3\rangle$
and $6P_{1/2},|F=4,m=-3\rangle$ are excited levels $U_{1}$and $U_{2}$,
respectively \cite{Julius2014superluminal}.  



\begin{thebibliography}{72}
\expandafter\ifx\csname natexlab\endcsname\relax\def\natexlab#1{#1}\fi
\expandafter\ifx\csname bibnamefont\endcsname\relax
  \def\bibnamefont#1{#1}\fi
\expandafter\ifx\csname bibfnamefont\endcsname\relax
  \def\bibfnamefont#1{#1}\fi
\expandafter\ifx\csname citenamefont\endcsname\relax
  \def\citenamefont#1{#1}\fi
\expandafter\ifx\csname url\endcsname\relax
  \def\url#1{\texttt{#1}}\fi
\expandafter\ifx\csname urlprefix\endcsname\relax\def\urlprefix{URL }\fi
\providecommand{\bibinfo}[2]{#2}
\providecommand{\eprint}[2][]{\url{#2}}

\bibitem[{\citenamefont{Harris}(1997)}]{Harris-EIT-1997}
\bibinfo{author}{\bibfnamefont{S.~E.} \bibnamefont{Harris}},
  \bibinfo{journal}{Physics Today} \textbf{\bibinfo{volume}{50}},
  \bibinfo{pages}{36} (\bibinfo{year}{1997}).

\bibitem[{\citenamefont{Fleischhauer et~al.}(2005)\citenamefont{Fleischhauer,
  Imamoglu, and Marangos}}]{Fleischhauer-RevModPhys-2005}
\bibinfo{author}{\bibfnamefont{M.}~\bibnamefont{Fleischhauer}},
  \bibinfo{author}{\bibfnamefont{A.}~\bibnamefont{Imamoglu}}, \bibnamefont{and}
  \bibinfo{author}{\bibfnamefont{J.~P.} \bibnamefont{Marangos}},
  \bibinfo{journal}{Rev. Mod. Phys.} \textbf{\bibinfo{volume}{77}},
  \bibinfo{pages}{633} (\bibinfo{year}{2005}),
  \urlprefix\url{https://link.aps.org/doi/10.1103/RevModPhys.77.633}.

\bibitem[{\citenamefont{Paspalakis et~al.}(2002)\citenamefont{Paspalakis,
  Kylstra, and Knight}}]{PaspalakisPhysRevA2002CPM}
\bibinfo{author}{\bibfnamefont{E.}~\bibnamefont{Paspalakis}},
  \bibinfo{author}{\bibfnamefont{N.~J.} \bibnamefont{Kylstra}},
  \bibnamefont{and} \bibinfo{author}{\bibfnamefont{P.~L.}
  \bibnamefont{Knight}}, \bibinfo{journal}{Phys. Rev. A}
  \textbf{\bibinfo{volume}{65}}, \bibinfo{pages}{053808}
  (\bibinfo{year}{2002}),
  \urlprefix\url{https://link.aps.org/doi/10.1103/PhysRevA.65.053808}.

\bibitem[{\citenamefont{Paspalakis and
  Kis}(2002{\natexlab{a}})}]{PaspalakisPhysRevA2002multi}
\bibinfo{author}{\bibfnamefont{E.}~\bibnamefont{Paspalakis}} \bibnamefont{and}
  \bibinfo{author}{\bibfnamefont{Z.}~\bibnamefont{Kis}},
  \bibinfo{journal}{Phys. Rev. A} \textbf{\bibinfo{volume}{66}},
  \bibinfo{pages}{025802} (\bibinfo{year}{2002}{\natexlab{a}}),
  \urlprefix\url{https://link.aps.org/doi/10.1103/PhysRevA.66.025802}.

\bibitem[{\citenamefont{Paspalakis and
  Kis}(2002{\natexlab{b}})}]{PaspalakisPRA2002}
\bibinfo{author}{\bibfnamefont{E.}~\bibnamefont{Paspalakis}} \bibnamefont{and}
  \bibinfo{author}{\bibfnamefont{Z.}~\bibnamefont{Kis}},
  \bibinfo{journal}{Phys. Rev. A} \textbf{\bibinfo{volume}{66}},
  \bibinfo{pages}{025802} (\bibinfo{year}{2002}{\natexlab{b}}),
  \urlprefix\url{https://link.aps.org/doi/10.1103/PhysRevA.66.025802}.

\bibitem[{\citenamefont{Ruseckas et~al.}(2007)\citenamefont{Ruseckas,
  Juzeli\=unas, \"Ohberg, and Barnett}}]{Ruseckas-PhysRevA-2007}
\bibinfo{author}{\bibfnamefont{J.}~\bibnamefont{Ruseckas}},
  \bibinfo{author}{\bibfnamefont{G.}~\bibnamefont{Juzeli\=unas}},
  \bibinfo{author}{\bibfnamefont{P.}~\bibnamefont{\"Ohberg}}, \bibnamefont{and}
  \bibinfo{author}{\bibfnamefont{S.~M.} \bibnamefont{Barnett}},
  \bibinfo{journal}{Phys. Rev. A} \textbf{\bibinfo{volume}{76}},
  \bibinfo{pages}{053822} (\bibinfo{year}{2007}),
  \urlprefix\url{https://link.aps.org/doi/10.1103/PhysRevA.76.053822}.

\bibitem[{\citenamefont{Grobe et~al.}(1994)\citenamefont{Grobe, Hioe, and
  Eberly}}]{Grobe-PhysRevLett-1994}
\bibinfo{author}{\bibfnamefont{R.}~\bibnamefont{Grobe}},
  \bibinfo{author}{\bibfnamefont{F.~T.} \bibnamefont{Hioe}}, \bibnamefont{and}
  \bibinfo{author}{\bibfnamefont{J.~H.} \bibnamefont{Eberly}},
  \bibinfo{journal}{Phys. Rev. Lett.} \textbf{\bibinfo{volume}{73}},
  \bibinfo{pages}{3183} (\bibinfo{year}{1994}),
  \urlprefix\url{https://link.aps.org/doi/10.1103/PhysRevLett.73.3183}.

\bibitem[{\citenamefont{Fleischhauer and
  Manka}(1996)}]{Fleischhauer-PhysRevA-1996}
\bibinfo{author}{\bibfnamefont{M.}~\bibnamefont{Fleischhauer}}
  \bibnamefont{and} \bibinfo{author}{\bibfnamefont{A.~S.} \bibnamefont{Manka}},
  \bibinfo{journal}{Phys. Rev. A} \textbf{\bibinfo{volume}{54}},
  \bibinfo{pages}{794} (\bibinfo{year}{1996}),
  \urlprefix\url{https://link.aps.org/doi/10.1103/PhysRevA.54.794}.

\bibitem[{\citenamefont{Wang et~al.}(2001)\citenamefont{Wang, Goorskey, and
  Xiao}}]{Wang-phys.rev.lett.2001}
\bibinfo{author}{\bibfnamefont{H.}~\bibnamefont{Wang}},
  \bibinfo{author}{\bibfnamefont{D.}~\bibnamefont{Goorskey}}, \bibnamefont{and}
  \bibinfo{author}{\bibfnamefont{M.}~\bibnamefont{Xiao}},
  \bibinfo{journal}{Phys. Rev. Lett.} \textbf{\bibinfo{volume}{87}},
  \bibinfo{pages}{073601} (\bibinfo{year}{2001}),
  \urlprefix\url{https://link.aps.org/doi/10.1103/PhysRevLett.87.073601}.

\bibitem[{\citenamefont{Harris}(1994)}]{Harris-PhysRevLett-1994}
\bibinfo{author}{\bibfnamefont{S.~E.} \bibnamefont{Harris}},
  \bibinfo{journal}{Phys. Rev. Lett.} \textbf{\bibinfo{volume}{72}},
  \bibinfo{pages}{52} (\bibinfo{year}{1994}),
  \urlprefix\url{https://link.aps.org/doi/10.1103/PhysRevLett.72.52}.

\bibitem[{\citenamefont{Cerboneschi and
  Arimondo}(1995)}]{Cerboneschi-PhysRevA-1995}
\bibinfo{author}{\bibfnamefont{E.}~\bibnamefont{Cerboneschi}} \bibnamefont{and}
  \bibinfo{author}{\bibfnamefont{E.}~\bibnamefont{Arimondo}},
  \bibinfo{journal}{Phys. Rev. A} \textbf{\bibinfo{volume}{52}},
  \bibinfo{pages}{R1823} (\bibinfo{year}{1995}),
  \urlprefix\url{https://link.aps.org/doi/10.1103/PhysRevA.52.R1823}.

\bibitem[{\citenamefont{Harris et~al.}(1990)\citenamefont{Harris, Field, and
  Imamo\ifmmode~\breve{g}\else \u{g}\fi{}lu}}]{Harris-PhysRevLett-1990}
\bibinfo{author}{\bibfnamefont{S.~E.} \bibnamefont{Harris}},
  \bibinfo{author}{\bibfnamefont{J.~E.} \bibnamefont{Field}}, \bibnamefont{and}
  \bibinfo{author}{\bibfnamefont{A.}~\bibnamefont{Imamo\ifmmode~\breve{g}\else
  \u{g}\fi{}lu}}, \bibinfo{journal}{Phys. Rev. Lett.}
  \textbf{\bibinfo{volume}{64}}, \bibinfo{pages}{1107} (\bibinfo{year}{1990}),
  \urlprefix\url{https://link.aps.org/doi/10.1103/PhysRevLett.64.1107}.

\bibitem[{\citenamefont{Deng et~al.}(1998)\citenamefont{Deng, Payne, and
  Garrett}}]{Deng-PhysRevA-1998}
\bibinfo{author}{\bibfnamefont{L.}~\bibnamefont{Deng}},
  \bibinfo{author}{\bibfnamefont{M.~G.} \bibnamefont{Payne}}, \bibnamefont{and}
  \bibinfo{author}{\bibfnamefont{W.~R.} \bibnamefont{Garrett}},
  \bibinfo{journal}{Phys. Rev. A} \textbf{\bibinfo{volume}{58}},
  \bibinfo{pages}{707} (\bibinfo{year}{1998}),
  \urlprefix\url{https://link.aps.org/doi/10.1103/PhysRevA.58.707}.

\bibitem[{\citenamefont{Kang and Zhu}(2003)}]{Kang-PhysRevLett-2003}
\bibinfo{author}{\bibfnamefont{H.}~\bibnamefont{Kang}} \bibnamefont{and}
  \bibinfo{author}{\bibfnamefont{Y.}~\bibnamefont{Zhu}},
  \bibinfo{journal}{Phys. Rev. Lett.} \textbf{\bibinfo{volume}{91}},
  \bibinfo{pages}{093601} (\bibinfo{year}{2003}),
  \urlprefix\url{https://link.aps.org/doi/10.1103/PhysRevLett.91.093601}.

\bibitem[{\citenamefont{Hamedi and Juzeli\ifmmode~\bar{u}\else
  \={u}\fi{}nas}(2015)}]{HamediPhysRevA.Kerr}
\bibinfo{author}{\bibfnamefont{H.~R.} \bibnamefont{Hamedi}} \bibnamefont{and}
  \bibinfo{author}{\bibfnamefont{G.}~\bibnamefont{Juzeli\ifmmode~\bar{u}\else
  \={u}\fi{}nas}}, \bibinfo{journal}{Phys. Rev. A}
  \textbf{\bibinfo{volume}{91}}, \bibinfo{pages}{053823}
  (\bibinfo{year}{2015}),
  \urlprefix\url{https://link.aps.org/doi/10.1103/PhysRevA.91.053823}.

\bibitem[{\citenamefont{Wu and Deng}(2004{\natexlab{a}})}]{Wu-PhysRevLett-2004}
\bibinfo{author}{\bibfnamefont{Y.}~\bibnamefont{Wu}} \bibnamefont{and}
  \bibinfo{author}{\bibfnamefont{L.}~\bibnamefont{Deng}},
  \bibinfo{journal}{Phys. Rev. Lett.} \textbf{\bibinfo{volume}{93}},
  \bibinfo{pages}{143904} (\bibinfo{year}{2004}{\natexlab{a}}),
  \urlprefix\url{https://link.aps.org/doi/10.1103/PhysRevLett.93.143904}.

\bibitem[{\citenamefont{Wu and Deng}(2004{\natexlab{b}})}]{Wu-OL-2004}
\bibinfo{author}{\bibfnamefont{Y.}~\bibnamefont{Wu}} \bibnamefont{and}
  \bibinfo{author}{\bibfnamefont{L.}~\bibnamefont{Deng}},
  \bibinfo{journal}{Opt. Lett.} \textbf{\bibinfo{volume}{29}},
  \bibinfo{pages}{2064} (\bibinfo{year}{2004}{\natexlab{b}}).

\bibitem[{\citenamefont{Hau et~al.}(1999)\citenamefont{Hau, Harris, Dutton, and
  Behroozi}}]{Lene1999slowlgiht}
\bibinfo{author}{\bibfnamefont{L.~V.} \bibnamefont{Hau}},
  \bibinfo{author}{\bibfnamefont{S.~E.} \bibnamefont{Harris}},
  \bibinfo{author}{\bibfnamefont{Z.}~\bibnamefont{Dutton}}, \bibnamefont{and}
  \bibinfo{author}{\bibfnamefont{C.~H.} \bibnamefont{Behroozi}},
  \bibinfo{journal}{Nature} \textbf{\bibinfo{volume}{397}},
  \bibinfo{pages}{594} (\bibinfo{year}{1999}).

\bibitem[{\citenamefont{Lukin}(2003)}]{Lukin-RevModPhys-2003}
\bibinfo{author}{\bibfnamefont{M.~D.} \bibnamefont{Lukin}},
  \bibinfo{journal}{Rev. Mod. Phys.} \textbf{\bibinfo{volume}{75}},
  \bibinfo{pages}{457} (\bibinfo{year}{2003}),
  \urlprefix\url{https://link.aps.org/doi/10.1103/RevModPhys.75.457}.

\bibitem[{\citenamefont{Juzeli\=unas and
  \"Ohberg}(2004)}]{Juzeliunas-PhysRevLett-2004}
\bibinfo{author}{\bibfnamefont{G.}~\bibnamefont{Juzeli\=unas}}
  \bibnamefont{and} \bibinfo{author}{\bibfnamefont{P.}~\bibnamefont{\"Ohberg}},
  \bibinfo{journal}{Phys. Rev. Lett.} \textbf{\bibinfo{volume}{93}},
  \bibinfo{pages}{033602} (\bibinfo{year}{2004}),
  \urlprefix\url{https://link.aps.org/doi/10.1103/PhysRevLett.93.033602}.

\bibitem[{\citenamefont{Ruseckas
  et~al.}(2011{\natexlab{a}})\citenamefont{Ruseckas, Kudria{\v{s}}ov,
  Juzeli{\=u}nas, Unanyan, Otterbach, and Fleischhauer}}]{Ruseckas2011PRA}
\bibinfo{author}{\bibfnamefont{J.}~\bibnamefont{Ruseckas}},
  \bibinfo{author}{\bibfnamefont{V.}~\bibnamefont{Kudria{\v{s}}ov}},
  \bibinfo{author}{\bibfnamefont{G.}~\bibnamefont{Juzeli{\=u}nas}},
  \bibinfo{author}{\bibfnamefont{R.~G.} \bibnamefont{Unanyan}},
  \bibinfo{author}{\bibfnamefont{J.}~\bibnamefont{Otterbach}},
  \bibnamefont{and}
  \bibinfo{author}{\bibfnamefont{M.}~\bibnamefont{Fleischhauer}},
  \bibinfo{journal}{Phys. Rev. A} \textbf{\bibinfo{volume}{83}},
  \bibinfo{pages}{063811} (\bibinfo{year}{2011}{\natexlab{a}}).

\bibitem[{\citenamefont{Bao et~al.}(2011)\citenamefont{Bao, Zhang, Gao, Zhang,
  Cui, and Wu}}]{Bao2011PRA}
\bibinfo{author}{\bibfnamefont{Q.-Q.} \bibnamefont{Bao}},
  \bibinfo{author}{\bibfnamefont{X.-H.} \bibnamefont{Zhang}},
  \bibinfo{author}{\bibfnamefont{J.-Y.} \bibnamefont{Gao}},
  \bibinfo{author}{\bibfnamefont{Y.}~\bibnamefont{Zhang}},
  \bibinfo{author}{\bibfnamefont{C.-L.} \bibnamefont{Cui}}, \bibnamefont{and}
  \bibinfo{author}{\bibfnamefont{J.-H.} \bibnamefont{Wu}},
  \bibinfo{journal}{Phys. Rev. A} \textbf{\bibinfo{volume}{84}},
  \bibinfo{pages}{063812} (\bibinfo{year}{2011}).

\bibitem[{\citenamefont{Ruseckas et~al.}(2013)\citenamefont{Ruseckas,
  Kudria\v{s}ov, Yu, and Juzeli\=unas}}]{Ruseckas-PhysRevA.87-2013}
\bibinfo{author}{\bibfnamefont{J.}~\bibnamefont{Ruseckas}},
  \bibinfo{author}{\bibfnamefont{V.}~\bibnamefont{Kudria\v{s}ov}},
  \bibinfo{author}{\bibfnamefont{I.~A.} \bibnamefont{Yu}}, \bibnamefont{and}
  \bibinfo{author}{\bibfnamefont{G.}~\bibnamefont{Juzeli\=unas}},
  \bibinfo{journal}{Phys. Rev. A} \textbf{\bibinfo{volume}{87}},
  \bibinfo{pages}{053840} (\bibinfo{year}{2013}),
  \urlprefix\url{https://link.aps.org/doi/10.1103/PhysRevA.87.053840}.

\bibitem[{\citenamefont{Lee et~al.}(2014)\citenamefont{Lee, Ruseckas, Lee,
  Kudria{\v{s}}ov, Chang, Cho, Juzeli{\=u}nas, and Yu}}]{Lee2014}
\bibinfo{author}{\bibfnamefont{M.-J.} \bibnamefont{Lee}},
  \bibinfo{author}{\bibfnamefont{J.}~\bibnamefont{Ruseckas}},
  \bibinfo{author}{\bibfnamefont{C.-Y.} \bibnamefont{Lee}},
  \bibinfo{author}{\bibfnamefont{V.}~\bibnamefont{Kudria{\v{s}}ov}},
  \bibinfo{author}{\bibfnamefont{K.-F.} \bibnamefont{Chang}},
  \bibinfo{author}{\bibfnamefont{H.-W.} \bibnamefont{Cho}},
  \bibinfo{author}{\bibfnamefont{G.}~\bibnamefont{Juzeli{\=u}nas}},
  \bibnamefont{and} \bibinfo{author}{\bibfnamefont{I.~A.} \bibnamefont{Yu}},
  \bibinfo{journal}{Nature Communications} \textbf{\bibinfo{volume}{5}},
  \bibinfo{pages}{5542} (\bibinfo{year}{2014}).

\bibitem[{\citenamefont{Phillips et~al.}(2001)\citenamefont{Phillips,
  Fleischhauer, Mair, Walsworth, and Lukin}}]{Phillips-PhysRevLett-2001}
\bibinfo{author}{\bibfnamefont{D.~F.} \bibnamefont{Phillips}},
  \bibinfo{author}{\bibfnamefont{A.}~\bibnamefont{Fleischhauer}},
  \bibinfo{author}{\bibfnamefont{A.}~\bibnamefont{Mair}},
  \bibinfo{author}{\bibfnamefont{R.~L.} \bibnamefont{Walsworth}},
  \bibnamefont{and} \bibinfo{author}{\bibfnamefont{M.~D.} \bibnamefont{Lukin}},
  \bibinfo{journal}{Phys. Rev. Lett.} \textbf{\bibinfo{volume}{86}},
  \bibinfo{pages}{783} (\bibinfo{year}{2001}),
  \urlprefix\url{https://link.aps.org/doi/10.1103/PhysRevLett.86.783}.

\bibitem[{\citenamefont{Liu et~al.}(2001)\citenamefont{Liu, Dutton, Behroozi,
  and Hau}}]{Chien-Nature-2001}
\bibinfo{author}{\bibfnamefont{C.}~\bibnamefont{Liu}},
  \bibinfo{author}{\bibfnamefont{Z.}~\bibnamefont{Dutton}},
  \bibinfo{author}{\bibfnamefont{C.~H.} \bibnamefont{Behroozi}},
  \bibnamefont{and} \bibinfo{author}{\bibfnamefont{L.~V.} \bibnamefont{Hau}},
  \bibinfo{journal}{Nature} \textbf{\bibinfo{volume}{409}}
  (\bibinfo{year}{2001}).

\bibitem[{\citenamefont{Lukin and Imamoglu}(2001)}]{Lukin-Nature-2001}
\bibinfo{author}{\bibfnamefont{M.~D.} \bibnamefont{Lukin}} \bibnamefont{and}
  \bibinfo{author}{\bibfnamefont{A.}~\bibnamefont{Imamoglu}},
  \bibinfo{journal}{Nature} \textbf{\bibinfo{volume}{413}}
  (\bibinfo{year}{2001}).

\bibitem[{\citenamefont{Juzeli\=unas and Carmichael}(2002)}]{Juzeliunas2002}
\bibinfo{author}{\bibfnamefont{G.}~\bibnamefont{Juzeli\=unas}}
  \bibnamefont{and} \bibinfo{author}{\bibfnamefont{H.~J.}
  \bibnamefont{Carmichael}}, \bibinfo{journal}{Phys. Rev. A}
  \textbf{\bibinfo{volume}{65}}, \bibinfo{pages}{021601(R)}
  (\bibinfo{year}{2002}).

\bibitem[{\citenamefont{Ma et~al.}(2017)\citenamefont{Ma, Slattery, and
  Tang}}]{Lijun-J.Opt-2017}
\bibinfo{author}{\bibfnamefont{L.}~\bibnamefont{Ma}},
  \bibinfo{author}{\bibfnamefont{O.}~\bibnamefont{Slattery}}, \bibnamefont{and}
  \bibinfo{author}{\bibfnamefont{X.}~\bibnamefont{Tang}}, \bibinfo{journal}{J.
  Opt.} \textbf{\bibinfo{volume}{19}}, \bibinfo{pages}{043001}
  (\bibinfo{year}{2017}).

\bibitem[{\citenamefont{Hsiao et~al.}(2018)\citenamefont{Hsiao, Tsai, Chen,
  Lin, Hung, Lee, Chen, Chen, Yu, and Chen}}]{Hsiao-PhysRevLett-2018}
\bibinfo{author}{\bibfnamefont{Y.-F.} \bibnamefont{Hsiao}},
  \bibinfo{author}{\bibfnamefont{P.-J.} \bibnamefont{Tsai}},
  \bibinfo{author}{\bibfnamefont{H.-S.} \bibnamefont{Chen}},
  \bibinfo{author}{\bibfnamefont{S.-X.} \bibnamefont{Lin}},
  \bibinfo{author}{\bibfnamefont{C.-C.} \bibnamefont{Hung}},
  \bibinfo{author}{\bibfnamefont{C.-H.} \bibnamefont{Lee}},
  \bibinfo{author}{\bibfnamefont{Y.-H.} \bibnamefont{Chen}},
  \bibinfo{author}{\bibfnamefont{Y.-F.} \bibnamefont{Chen}},
  \bibinfo{author}{\bibfnamefont{I.~A.} \bibnamefont{Yu}}, \bibnamefont{and}
  \bibinfo{author}{\bibfnamefont{Y.-C.} \bibnamefont{Chen}},
  \bibinfo{journal}{Phys. Rev. Lett.} \textbf{\bibinfo{volume}{120}},
  \bibinfo{pages}{183602} (\bibinfo{year}{2018}),
  \urlprefix\url{https://link.aps.org/doi/10.1103/PhysRevLett.120.183602}.

\bibitem[{\citenamefont{Jiang et~al.}(2007)\citenamefont{Jiang, Deng, and
  Payne}}]{Jiang2007}
\bibinfo{author}{\bibfnamefont{K.~J.} \bibnamefont{Jiang}},
  \bibinfo{author}{\bibfnamefont{L.}~\bibnamefont{Deng}}, \bibnamefont{and}
  \bibinfo{author}{\bibfnamefont{M.~G.} \bibnamefont{Payne}},
  \bibinfo{journal}{Phys. Rev. A} \textbf{\bibinfo{volume}{76}},
  \bibinfo{pages}{033819} (\bibinfo{year}{2007}),
  \urlprefix\url{https://link.aps.org/doi/10.1103/PhysRevA.76.033819}.

\bibitem[{\citenamefont{M.Mahmoudi et~al.}(2006)\citenamefont{M.Mahmoudi,
  M.Sahrai, and H.Tajalli}}]{Mahmoudi2006}
\bibinfo{author}{\bibnamefont{M.Mahmoudi}},
  \bibinfo{author}{\bibnamefont{M.Sahrai}}, \bibnamefont{and}
  \bibinfo{author}{\bibnamefont{H.Tajalli}}, \bibinfo{journal}{Phys. Lett. A}
  \textbf{\bibinfo{volume}{357}}, \bibinfo{pages}{66} (\bibinfo{year}{2006}).

\bibitem[{\citenamefont{qi~Kuang et~al.}(2009)\citenamefont{qi~Kuang, gang Wan,
  Kou, Jiang, , and yue Gao}}]{Shang2009}
\bibinfo{author}{\bibfnamefont{S.}~\bibnamefont{qi~Kuang}},
  \bibinfo{author}{\bibfnamefont{R.}~\bibnamefont{gang Wan}},
  \bibinfo{author}{\bibfnamefont{J.}~\bibnamefont{Kou}},
  \bibinfo{author}{\bibfnamefont{Y.}~\bibnamefont{Jiang}}, , \bibnamefont{and}
  \bibinfo{author}{\bibfnamefont{J.}~\bibnamefont{yue Gao}},
  \bibinfo{journal}{J. Opt. Soc. Am. B} \textbf{\bibinfo{volume}{26}},
  \bibinfo{pages}{2256} (\bibinfo{year}{2009}).

\bibitem[{\citenamefont{Akulshin and McLean}(2010)}]{AlexanderJOptics2010}
\bibinfo{author}{\bibfnamefont{A.~M.} \bibnamefont{Akulshin}} \bibnamefont{and}
  \bibinfo{author}{\bibfnamefont{R.~J.} \bibnamefont{McLean}},
  \bibinfo{journal}{J. Opt.} \textbf{\bibinfo{volume}{12}},
  \bibinfo{pages}{104001} (\bibinfo{year}{2010}).

\bibitem[{\citenamefont{Chu and Wong}(1982)}]{Chu1982}
\bibinfo{author}{\bibfnamefont{S.}~\bibnamefont{Chu}} \bibnamefont{and}
  \bibinfo{author}{\bibfnamefont{S.}~\bibnamefont{Wong}},
  \bibinfo{journal}{Phys. Rev. Lett.} \textbf{\bibinfo{volume}{48}},
  \bibinfo{pages}{738} (\bibinfo{year}{1982}),
  \urlprefix\url{https://link.aps.org/doi/10.1103/PhysRevLett.48.738}.

\bibitem[{\citenamefont{Steinberg et~al.}(1993)\citenamefont{Steinberg, Kwiat,
  and Chiao}}]{Steinberg1993}
\bibinfo{author}{\bibfnamefont{A.~M.} \bibnamefont{Steinberg}},
  \bibinfo{author}{\bibfnamefont{P.~G.} \bibnamefont{Kwiat}}, \bibnamefont{and}
  \bibinfo{author}{\bibfnamefont{R.~Y.} \bibnamefont{Chiao}},
  \bibinfo{journal}{Phys. Rev. Lett.} \textbf{\bibinfo{volume}{71}},
  \bibinfo{pages}{708} (\bibinfo{year}{1993}),
  \urlprefix\url{https://link.aps.org/doi/10.1103/PhysRevLett.71.708}.

\bibitem[{\citenamefont{Chiao}(1993)}]{Chiao1993}
\bibinfo{author}{\bibfnamefont{R.~Y.} \bibnamefont{Chiao}},
  \bibinfo{journal}{Phys. Rev. A} \textbf{\bibinfo{volume}{48}},
  \bibinfo{pages}{R34} (\bibinfo{year}{1993}),
  \urlprefix\url{https://link.aps.org/doi/10.1103/PhysRevA.48.R34}.

\bibitem[{\citenamefont{Dogariu et~al.}(2001)\citenamefont{Dogariu, Kuzmich,
  and Wang}}]{Dogariu2001}
\bibinfo{author}{\bibfnamefont{A.}~\bibnamefont{Dogariu}},
  \bibinfo{author}{\bibfnamefont{A.}~\bibnamefont{Kuzmich}}, \bibnamefont{and}
  \bibinfo{author}{\bibfnamefont{L.~J.} \bibnamefont{Wang}},
  \bibinfo{journal}{Phys. Rev. A} \textbf{\bibinfo{volume}{63}},
  \bibinfo{pages}{053806} (\bibinfo{year}{2001}),
  \urlprefix\url{https://link.aps.org/doi/10.1103/PhysRevA.63.053806}.

\bibitem[{\citenamefont{Glasser et~al.}(2012)\citenamefont{Glasser, Vogl, and
  Lett}}]{Glasser2012}
\bibinfo{author}{\bibfnamefont{R.~T.} \bibnamefont{Glasser}},
  \bibinfo{author}{\bibfnamefont{U.}~\bibnamefont{Vogl}}, \bibnamefont{and}
  \bibinfo{author}{\bibfnamefont{P.~D.} \bibnamefont{Lett}},
  \bibinfo{journal}{Phys. Rev. Lett.} \textbf{\bibinfo{volume}{108}},
  \bibinfo{pages}{173902} (\bibinfo{year}{2012}),
  \urlprefix\url{https://link.aps.org/doi/10.1103/PhysRevLett.108.173902}.

\bibitem[{\citenamefont{Bianucci et~al.}(2008)\citenamefont{Bianucci, Fietz,
  Robertson, Shvets, and Shih}}]{Bianucci2008}
\bibinfo{author}{\bibfnamefont{P.}~\bibnamefont{Bianucci}},
  \bibinfo{author}{\bibfnamefont{C.~R.} \bibnamefont{Fietz}},
  \bibinfo{author}{\bibfnamefont{J.~W.} \bibnamefont{Robertson}},
  \bibinfo{author}{\bibfnamefont{G.}~\bibnamefont{Shvets}}, \bibnamefont{and}
  \bibinfo{author}{\bibfnamefont{C.-K.} \bibnamefont{Shih}},
  \bibinfo{journal}{Phys. Rev. A} \textbf{\bibinfo{volume}{77}},
  \bibinfo{pages}{053816} (\bibinfo{year}{2008}),
  \urlprefix\url{https://link.aps.org/doi/10.1103/PhysRevA.77.053816}.

\bibitem[{\citenamefont{Kudria\ifmmode~\check{s}\else \v{s}\fi{}ov
  et~al.}(2014)\citenamefont{Kudria\ifmmode~\check{s}\else \v{s}\fi{}ov,
  Ruseckas, Mekys, Ekers, Bezuglov, and Juzeli\ifmmode~\bar{u}\else
  \={u}\fi{}nas}}]{Julius2014superluminal}
\bibinfo{author}{\bibfnamefont{V.}~\bibnamefont{Kudria\ifmmode~\check{s}\else
  \v{s}\fi{}ov}}, \bibinfo{author}{\bibfnamefont{J.}~\bibnamefont{Ruseckas}},
  \bibinfo{author}{\bibfnamefont{A.}~\bibnamefont{Mekys}},
  \bibinfo{author}{\bibfnamefont{A.}~\bibnamefont{Ekers}},
  \bibinfo{author}{\bibfnamefont{N.}~\bibnamefont{Bezuglov}}, \bibnamefont{and}
  \bibinfo{author}{\bibfnamefont{G.}~\bibnamefont{Juzeli\ifmmode~\bar{u}\else
  \={u}\fi{}nas}}, \bibinfo{journal}{Phys. Rev. A}
  \textbf{\bibinfo{volume}{90}}, \bibinfo{pages}{033827}
  (\bibinfo{year}{2014}),
  \urlprefix\url{https://link.aps.org/doi/10.1103/PhysRevA.90.033827}.

\bibitem[{\citenamefont{Allen et~al.}(1999)\citenamefont{Allen, Padgett, and
  Babiker}}]{Allen1999}
\bibinfo{author}{\bibfnamefont{L.}~\bibnamefont{Allen}},
  \bibinfo{author}{\bibfnamefont{M.~J.} \bibnamefont{Padgett}},
  \bibnamefont{and} \bibinfo{author}{\bibfnamefont{M.}~\bibnamefont{Babiker}},
  \bibinfo{journal}{Progress in Optics} \textbf{\bibinfo{volume}{39}},
  \bibinfo{pages}{291} (\bibinfo{year}{1999}).

\bibitem[{\citenamefont{Padgett et~al.}(2004)\citenamefont{Padgett, Courtial,
  and Allen}}]{Miles-physToday-2004}
\bibinfo{author}{\bibfnamefont{M.}~\bibnamefont{Padgett}},
  \bibinfo{author}{\bibfnamefont{J.}~\bibnamefont{Courtial}}, \bibnamefont{and}
  \bibinfo{author}{\bibfnamefont{L.}~\bibnamefont{Allen}},
  \bibinfo{journal}{Physics Today} \textbf{\bibinfo{volume}{57}},
  \bibinfo{pages}{35} (\bibinfo{year}{2004}).

\bibitem[{\citenamefont{Babiker et~al.}(2018)\citenamefont{Babiker, Andrews,
  and Lembessis}}]{Babiker2018}
\bibinfo{author}{\bibfnamefont{M.}~\bibnamefont{Babiker}},
  \bibinfo{author}{\bibfnamefont{D.~L.} \bibnamefont{Andrews}},
  \bibnamefont{and} \bibinfo{author}{\bibfnamefont{V.~E.}
  \bibnamefont{Lembessis}}, \bibinfo{journal}{Journal of Optics}
  \textbf{\bibinfo{volume}{21}}, \bibinfo{pages}{013001}
  (\bibinfo{year}{2018}),
  \urlprefix\url{https://doi.org/10.1088%2F2040-8986%2Faaed14}.

\bibitem[{\citenamefont{Babiker et~al.}(1994)\citenamefont{Babiker, Power, and
  Allen}}]{Babiker-PhysRevLett1994}
\bibinfo{author}{\bibfnamefont{M.}~\bibnamefont{Babiker}},
  \bibinfo{author}{\bibfnamefont{W.~L.} \bibnamefont{Power}}, \bibnamefont{and}
  \bibinfo{author}{\bibfnamefont{L.}~\bibnamefont{Allen}},
  \bibinfo{journal}{Phys. Rev. Lett.} \textbf{\bibinfo{volume}{73}},
  \bibinfo{pages}{1239} (\bibinfo{year}{1994}),
  \urlprefix\url{https://link.aps.org/doi/10.1103/PhysRevLett.73.1239}.

\bibitem[{\citenamefont{Molina-Terriza
  et~al.}(2001)\citenamefont{Molina-Terriza, Torres, and Torner}}]{Molina2001}
\bibinfo{author}{\bibfnamefont{G.}~\bibnamefont{Molina-Terriza}},
  \bibinfo{author}{\bibfnamefont{J.~P.} \bibnamefont{Torres}},
  \bibnamefont{and} \bibinfo{author}{\bibfnamefont{L.}~\bibnamefont{Torner}},
  \bibinfo{journal}{Phys. Rev. Lett.} \textbf{\bibinfo{volume}{88}},
  \bibinfo{pages}{013601} (\bibinfo{year}{2001}),
  \urlprefix\url{https://link.aps.org/doi/10.1103/PhysRevLett.88.013601}.

\bibitem[{\citenamefont{Pugatch et~al.}(2007)\citenamefont{Pugatch, Shuker,
  Firstenberg, Ron, and Davidson}}]{Pugatch-PhysRevLet-2004}
\bibinfo{author}{\bibfnamefont{R.}~\bibnamefont{Pugatch}},
  \bibinfo{author}{\bibfnamefont{M.}~\bibnamefont{Shuker}},
  \bibinfo{author}{\bibfnamefont{O.}~\bibnamefont{Firstenberg}},
  \bibinfo{author}{\bibfnamefont{A.}~\bibnamefont{Ron}}, \bibnamefont{and}
  \bibinfo{author}{\bibfnamefont{N.}~\bibnamefont{Davidson}},
  \bibinfo{journal}{Phys. Rev. Lett.} \textbf{\bibinfo{volume}{98}},
  \bibinfo{pages}{203601} (\bibinfo{year}{2007}),
  \urlprefix\url{https://link.aps.org/doi/10.1103/PhysRevLett.98.203601}.

\bibitem[{\citenamefont{Dutton and
  Ruostekoski}(2004)}]{Dutton-PhysRevLett-2004}
\bibinfo{author}{\bibfnamefont{Z.}~\bibnamefont{Dutton}} \bibnamefont{and}
  \bibinfo{author}{\bibfnamefont{J.}~\bibnamefont{Ruostekoski}},
  \bibinfo{journal}{Phys. Rev. Lett.} \textbf{\bibinfo{volume}{93}},
  \bibinfo{pages}{193602} (\bibinfo{year}{2004}),
  \urlprefix\url{https://link.aps.org/doi/10.1103/PhysRevLett.93.193602}.

\bibitem[{\citenamefont{Bishop et~al.}(2004)\citenamefont{Bishop, Nieminen,
  Heckenberg, and Rubinsztein-Dunlop}}]{Bishop2004}
\bibinfo{author}{\bibfnamefont{A.~I.} \bibnamefont{Bishop}},
  \bibinfo{author}{\bibfnamefont{T.~A.} \bibnamefont{Nieminen}},
  \bibinfo{author}{\bibfnamefont{N.~R.} \bibnamefont{Heckenberg}},
  \bibnamefont{and}
  \bibinfo{author}{\bibfnamefont{H.}~\bibnamefont{Rubinsztein-Dunlop}},
  \bibinfo{journal}{Phys. Rev. Lett.} \textbf{\bibinfo{volume}{92}},
  \bibinfo{pages}{198104} (\bibinfo{year}{2004}),
  \urlprefix\url{https://link.aps.org/doi/10.1103/PhysRevLett.92.198104}.

\bibitem[{\citenamefont{Chen et~al.}(2008)\citenamefont{Chen, Shi, Zhang, and
  Guo}}]{Chen-PhysRevA-2008}
\bibinfo{author}{\bibfnamefont{Q.-F.} \bibnamefont{Chen}},
  \bibinfo{author}{\bibfnamefont{B.-S.} \bibnamefont{Shi}},
  \bibinfo{author}{\bibfnamefont{Y.-S.} \bibnamefont{Zhang}}, \bibnamefont{and}
  \bibinfo{author}{\bibfnamefont{G.-C.} \bibnamefont{Guo}},
  \bibinfo{journal}{Phys. Rev. A} \textbf{\bibinfo{volume}{78}},
  \bibinfo{pages}{053810} (\bibinfo{year}{2008}),
  \urlprefix\url{https://link.aps.org/doi/10.1103/PhysRevA.78.053810}.

\bibitem[{\citenamefont{Lembessis and Babiker}(2010)}]{Lembessis-PhysRevA-2010}
\bibinfo{author}{\bibfnamefont{V.~E.} \bibnamefont{Lembessis}}
  \bibnamefont{and} \bibinfo{author}{\bibfnamefont{M.}~\bibnamefont{Babiker}},
  \bibinfo{journal}{Phys. Rev. A} \textbf{\bibinfo{volume}{82}},
  \bibinfo{pages}{051402} (\bibinfo{year}{2010}),
  \urlprefix\url{https://link.aps.org/doi/10.1103/PhysRevA.82.051402}.

\bibitem[{\citenamefont{Ruseckas
  et~al.}(2011{\natexlab{b}})\citenamefont{Ruseckas, Mekys, and
  Juzeli\=unas}}]{Ruseckas2011}
\bibinfo{author}{\bibfnamefont{J.}~\bibnamefont{Ruseckas}},
  \bibinfo{author}{\bibfnamefont{A.}~\bibnamefont{Mekys}}, \bibnamefont{and}
  \bibinfo{author}{\bibfnamefont{G.}~\bibnamefont{Juzeli\=unas}},
  \bibinfo{journal}{J. Opt.} \textbf{\bibinfo{volume}{13}},
  \bibinfo{pages}{064013} (\bibinfo{year}{2011}{\natexlab{b}}).

\bibitem[{\citenamefont{Ding et~al.}(2012)\citenamefont{Ding, Zhou, Shi, Zou,
  and Guo}}]{Ding-OL-2012}
\bibinfo{author}{\bibfnamefont{D.-S.} \bibnamefont{Ding}},
  \bibinfo{author}{\bibfnamefont{Z.-Y.} \bibnamefont{Zhou}},
  \bibinfo{author}{\bibfnamefont{B.-S.} \bibnamefont{Shi}},
  \bibinfo{author}{\bibfnamefont{X.-B.} \bibnamefont{Zou}}, \bibnamefont{and}
  \bibinfo{author}{\bibfnamefont{G.-C.} \bibnamefont{Guo}},
  \bibinfo{journal}{Opt. Lett.} \textbf{\bibinfo{volume}{37}},
  \bibinfo{pages}{3270} (\bibinfo{year}{2012}).

\bibitem[{\citenamefont{Walker et~al.}(2012)\citenamefont{Walker, Arnold, and
  Franke-Arnold}}]{WalkerPhysRevLett2012}
\bibinfo{author}{\bibfnamefont{G.}~\bibnamefont{Walker}},
  \bibinfo{author}{\bibfnamefont{A.~S.} \bibnamefont{Arnold}},
  \bibnamefont{and}
  \bibinfo{author}{\bibfnamefont{S.}~\bibnamefont{Franke-Arnold}},
  \bibinfo{journal}{Phys. Rev. Lett.} \textbf{\bibinfo{volume}{108}},
  \bibinfo{pages}{243601} (\bibinfo{year}{2012}),
  \urlprefix\url{https://link.aps.org/doi/10.1103/PhysRevLett.108.243601}.

\bibitem[{\citenamefont{Lembessis et~al.}(2014)\citenamefont{Lembessis,
  Ellinas, Babiker, and Al-Dossary}}]{Lembessis-PhysRevA.89-2014}
\bibinfo{author}{\bibfnamefont{V.~E.} \bibnamefont{Lembessis}},
  \bibinfo{author}{\bibfnamefont{D.}~\bibnamefont{Ellinas}},
  \bibinfo{author}{\bibfnamefont{M.}~\bibnamefont{Babiker}}, \bibnamefont{and}
  \bibinfo{author}{\bibfnamefont{O.}~\bibnamefont{Al-Dossary}},
  \bibinfo{journal}{Phys. Rev. A} \textbf{\bibinfo{volume}{89}},
  \bibinfo{pages}{053616} (\bibinfo{year}{2014}),
  \urlprefix\url{https://link.aps.org/doi/10.1103/PhysRevA.89.053616}.

\bibitem[{\citenamefont{Radwell et~al.}(2015)\citenamefont{Radwell, Clark,
  Piccirillo, Barnett, and Franke-Arnold}}]{RadwellPhysRevLet2015}
\bibinfo{author}{\bibfnamefont{N.}~\bibnamefont{Radwell}},
  \bibinfo{author}{\bibfnamefont{T.~W.} \bibnamefont{Clark}},
  \bibinfo{author}{\bibfnamefont{B.}~\bibnamefont{Piccirillo}},
  \bibinfo{author}{\bibfnamefont{S.~M.} \bibnamefont{Barnett}},
  \bibnamefont{and}
  \bibinfo{author}{\bibfnamefont{S.}~\bibnamefont{Franke-Arnold}},
  \bibinfo{journal}{Phys. Rev. Lett.} \textbf{\bibinfo{volume}{114}},
  \bibinfo{pages}{123603} (\bibinfo{year}{2015}),
  \urlprefix\url{https://link.aps.org/doi/10.1103/PhysRevLett.114.123603}.

\bibitem[{\citenamefont{Sharma and Dey}(2017)}]{SharmaPhysRevA2017}
\bibinfo{author}{\bibfnamefont{S.}~\bibnamefont{Sharma}} \bibnamefont{and}
  \bibinfo{author}{\bibfnamefont{T.~N.} \bibnamefont{Dey}},
  \bibinfo{journal}{Phys. Rev. A} \textbf{\bibinfo{volume}{96}},
  \bibinfo{pages}{033811} (\bibinfo{year}{2017}),
  \urlprefix\url{https://link.aps.org/doi/10.1103/PhysRevA.96.033811}.

\bibitem[{\citenamefont{Hamedi et~al.}(2018{\natexlab{a}})\citenamefont{Hamedi,
  Kudria\v{s}ov, Ruseckas, and Juzeli\=unas}}]{Hamedi2018OE}
\bibinfo{author}{\bibfnamefont{H.~R.} \bibnamefont{Hamedi}},
  \bibinfo{author}{\bibfnamefont{V.}~\bibnamefont{Kudria\v{s}ov}},
  \bibinfo{author}{\bibfnamefont{J.}~\bibnamefont{Ruseckas}}, \bibnamefont{and}
  \bibinfo{author}{\bibfnamefont{G.}~\bibnamefont{Juzeli\=unas}},
  \bibinfo{journal}{Optics Express} \textbf{\bibinfo{volume}{26}},
  \bibinfo{pages}{28249} (\bibinfo{year}{2018}{\natexlab{a}}).

\bibitem[{\citenamefont{Hamedi et~al.}(2018{\natexlab{b}})\citenamefont{Hamedi,
  Ruseckas, and Juzeli\=unas}}]{Hamedi-PhysRevA-2018}
\bibinfo{author}{\bibfnamefont{H.~R.} \bibnamefont{Hamedi}},
  \bibinfo{author}{\bibfnamefont{J.}~\bibnamefont{Ruseckas}}, \bibnamefont{and}
  \bibinfo{author}{\bibfnamefont{G.}~\bibnamefont{Juzeli\=unas}},
  \bibinfo{journal}{Phys. Rev. A} \textbf{\bibinfo{volume}{98}},
  \bibinfo{pages}{013840} (\bibinfo{year}{2018}{\natexlab{b}}),
  \urlprefix\url{https://link.aps.org/doi/10.1103/PhysRevA.98.013840}.

\bibitem[{\citenamefont{Hamedi et~al.}(2019{\natexlab{a}})\citenamefont{Hamedi,
  Ruseckas, Paspalakis, and Juzeli\ifmmode~\bar{u}\else
  \={u}\fi{}nas}}]{Hamedipra2019}
\bibinfo{author}{\bibfnamefont{H.~R.} \bibnamefont{Hamedi}},
  \bibinfo{author}{\bibfnamefont{J.}~\bibnamefont{Ruseckas}},
  \bibinfo{author}{\bibfnamefont{E.}~\bibnamefont{Paspalakis}},
  \bibnamefont{and}
  \bibinfo{author}{\bibfnamefont{G.}~\bibnamefont{Juzeli\ifmmode~\bar{u}\else
  \={u}\fi{}nas}}, \bibinfo{journal}{Phys. Rev. A}
  \textbf{\bibinfo{volume}{99}}, \bibinfo{pages}{033812}
  (\bibinfo{year}{2019}{\natexlab{a}}),
  \urlprefix\url{https://link.aps.org/doi/10.1103/PhysRevA.99.033812}.

\bibitem[{\citenamefont{Moretti et~al.}(2009)\citenamefont{Moretti, Felinto,
  and Tabosa}}]{Moretti-PhysRevA-2009}
\bibinfo{author}{\bibfnamefont{D.}~\bibnamefont{Moretti}},
  \bibinfo{author}{\bibfnamefont{D.}~\bibnamefont{Felinto}}, \bibnamefont{and}
  \bibinfo{author}{\bibfnamefont{J.~W.~R.} \bibnamefont{Tabosa}},
  \bibinfo{journal}{Phys. Rev. A} \textbf{\bibinfo{volume}{79}},
  \bibinfo{pages}{023825} (\bibinfo{year}{2009}),
  \urlprefix\url{https://link.aps.org/doi/10.1103/PhysRevA.79.023825}.

\bibitem[{\citenamefont{Cardano et~al.}(2015)\citenamefont{Cardano, Massa,
  Qassim, Karimi, Slussarenko, Paparo, de~Lisio, Sciarrino, Santamato, Boyd
  et~al.}}]{Filippo2015}
\bibinfo{author}{\bibfnamefont{F.}~\bibnamefont{Cardano}},
  \bibinfo{author}{\bibfnamefont{F.}~\bibnamefont{Massa}},
  \bibinfo{author}{\bibfnamefont{H.}~\bibnamefont{Qassim}},
  \bibinfo{author}{\bibfnamefont{E.}~\bibnamefont{Karimi}},
  \bibinfo{author}{\bibfnamefont{S.}~\bibnamefont{Slussarenko}},
  \bibinfo{author}{\bibfnamefont{D.}~\bibnamefont{Paparo}},
  \bibinfo{author}{\bibfnamefont{C.}~\bibnamefont{de~Lisio}},
  \bibinfo{author}{\bibfnamefont{F.}~\bibnamefont{Sciarrino}},
  \bibinfo{author}{\bibfnamefont{E.}~\bibnamefont{Santamato}},
  \bibinfo{author}{\bibfnamefont{R.~W.} \bibnamefont{Boyd}},
  \bibnamefont{et~al.}, \bibinfo{journal}{Science Advances}
  \textbf{\bibinfo{volume}{1}}, \bibinfo{pages}{e1500087}
  (\bibinfo{year}{2015}).

\bibitem[{\citenamefont{Hamedi et~al.}(2019{\natexlab{b}})\citenamefont{Hamedi,
  Paspalakis, \ifmmode~\check{Z}\else \v{Z}\fi{}labys,
  Juzeli\ifmmode~\bar{u}\else \={u}\fi{}nas, and Ruseckas}}]{HamidCPT2019}
\bibinfo{author}{\bibfnamefont{H.~R.} \bibnamefont{Hamedi}},
  \bibinfo{author}{\bibfnamefont{E.}~\bibnamefont{Paspalakis}},
  \bibinfo{author}{\bibfnamefont{G.}~\bibnamefont{\ifmmode~\check{Z}\else
  \v{Z}\fi{}labys}},
  \bibinfo{author}{\bibfnamefont{G.}~\bibnamefont{Juzeli\ifmmode~\bar{u}\else
  \={u}\fi{}nas}}, \bibnamefont{and}
  \bibinfo{author}{\bibfnamefont{J.}~\bibnamefont{Ruseckas}},
  \bibinfo{journal}{Phys. Rev. A} \textbf{\bibinfo{volume}{100}},
  \bibinfo{pages}{023811} (\bibinfo{year}{2019}{\natexlab{b}}),
  \urlprefix\url{https://link.aps.org/doi/10.1103/PhysRevA.100.023811}.

\bibitem[{\citenamefont{Hong et~al.}(2019)\citenamefont{Hong, Wang, Ding, and
  Yu}}]{Yin2019FWM}
\bibinfo{author}{\bibfnamefont{Y.}~\bibnamefont{Hong}},
  \bibinfo{author}{\bibfnamefont{Z.}~\bibnamefont{Wang}},
  \bibinfo{author}{\bibfnamefont{D.}~\bibnamefont{Ding}}, \bibnamefont{and}
  \bibinfo{author}{\bibfnamefont{B.}~\bibnamefont{Yu}},
  \bibinfo{journal}{Optics Express} \textbf{\bibinfo{volume}{27}},
  \bibinfo{pages}{29863} (\bibinfo{year}{2019}).

\bibitem[{\citenamefont{Qiu et~al.}(2020)\citenamefont{Qiu, Wang, Ding, Li, ,
  and Yu}}]{Jing2020FWM}
\bibinfo{author}{\bibfnamefont{J.}~\bibnamefont{Qiu}},
  \bibinfo{author}{\bibfnamefont{Z.}~\bibnamefont{Wang}},
  \bibinfo{author}{\bibfnamefont{D.}~\bibnamefont{Ding}},
  \bibinfo{author}{\bibfnamefont{W.}~\bibnamefont{Li}}, , \bibnamefont{and}
  \bibinfo{author}{\bibfnamefont{B.}~\bibnamefont{Yu}},
  \bibinfo{journal}{Optics Express} \textbf{\bibinfo{volume}{28}},
  \bibinfo{pages}{2975} (\bibinfo{year}{2020}).

\bibitem[{\citenamefont{Maleev and Swartzlander}(2003)}]{Ivan2003}
\bibinfo{author}{\bibfnamefont{I.~D.} \bibnamefont{Maleev}} \bibnamefont{and}
  \bibinfo{author}{\bibfnamefont{G.~A.} \bibnamefont{Swartzlander}},
  \bibinfo{journal}{J. Opt. Soc. Am. B} \textbf{\bibinfo{volume}{20}},
  \bibinfo{pages}{1169} (\bibinfo{year}{2003}).

\bibitem[{\citenamefont{Galvez et~al.}(2006)\citenamefont{Galvez, Smiley, and
  Fernandes}}]{Galvez2006}
\bibinfo{author}{\bibfnamefont{E.~J.} \bibnamefont{Galvez}},
  \bibinfo{author}{\bibfnamefont{N.}~\bibnamefont{Smiley}}, \bibnamefont{and}
  \bibinfo{author}{\bibfnamefont{N.}~\bibnamefont{Fernandes}},
  \bibinfo{journal}{Proc. SPIE} \textbf{\bibinfo{volume}{6131}},
  \bibinfo{pages}{613105} (\bibinfo{year}{2006}).

\bibitem[{\citenamefont{Baumann et~al.}(2009)\citenamefont{Baumann, Kalb,
  MacMillan, and Galvez}}]{Baumann2009}
\bibinfo{author}{\bibfnamefont{S.~M.} \bibnamefont{Baumann}},
  \bibinfo{author}{\bibfnamefont{D.~M.} \bibnamefont{Kalb}},
  \bibinfo{author}{\bibfnamefont{L.~H.} \bibnamefont{MacMillan}},
  \bibnamefont{and} \bibinfo{author}{\bibfnamefont{E.~J.}
  \bibnamefont{Galvez}}, \bibinfo{journal}{Optics Express}
  \textbf{\bibinfo{volume}{17}}, \bibinfo{pages}{9818} (\bibinfo{year}{2009}).

\bibitem[{\citenamefont{Mair et~al.}(2001)\citenamefont{Mair, Vaziri, Weihs,
  and Zeilinger}}]{Alois2001}
\bibinfo{author}{\bibfnamefont{A.}~\bibnamefont{Mair}},
  \bibinfo{author}{\bibfnamefont{A.}~\bibnamefont{Vaziri}},
  \bibinfo{author}{\bibfnamefont{G.}~\bibnamefont{Weihs}}, \bibnamefont{and}
  \bibinfo{author}{\bibfnamefont{A.}~\bibnamefont{Zeilinger}},
  \bibinfo{journal}{Nature} \textbf{\bibinfo{volume}{412}}
  (\bibinfo{year}{2001}).

\bibitem[{\citenamefont{Basistiy et~al.}(2003)\citenamefont{Basistiy, Slyusar,
  Soskin, Vasnetsov, and Bekshaev}}]{Basistiy2003}
\bibinfo{author}{\bibfnamefont{I.~V.} \bibnamefont{Basistiy}},
  \bibinfo{author}{\bibfnamefont{V.~V.} \bibnamefont{Slyusar}},
  \bibinfo{author}{\bibfnamefont{M.~S.} \bibnamefont{Soskin}},
  \bibinfo{author}{\bibfnamefont{M.~V.} \bibnamefont{Vasnetsov}},
  \bibnamefont{and} \bibinfo{author}{\bibfnamefont{A.~Y.}
  \bibnamefont{Bekshaev}}, \bibinfo{journal}{Opt. Lett.}
  \textbf{\bibinfo{volume}{28}}, \bibinfo{pages}{1185} (\bibinfo{year}{2003}).

\bibitem[{\citenamefont{Lee et~al.}(2005)\citenamefont{Lee, Ahluwalia, Yuan,
  Cheong, and Dholakia}}]{Lee2005}
\bibinfo{author}{\bibfnamefont{W.~M.} \bibnamefont{Lee}},
  \bibinfo{author}{\bibfnamefont{B.~P.~S.} \bibnamefont{Ahluwalia}},
  \bibinfo{author}{\bibfnamefont{X.-C.} \bibnamefont{Yuan}},
  \bibinfo{author}{\bibfnamefont{W.~C.} \bibnamefont{Cheong}},
  \bibnamefont{and} \bibinfo{author}{\bibfnamefont{K.}~\bibnamefont{Dholakia}},
  \bibinfo{journal}{J. Opt. A: Pure Appl. Opt.} \textbf{\bibinfo{volume}{7}},
  \bibinfo{pages}{1} (\bibinfo{year}{2005}).

\bibitem[{\citenamefont{Wang}(2017)}]{Zhang2017}
\bibinfo{author}{\bibfnamefont{X.~Z.~H.} \bibnamefont{Wang}},
  \bibinfo{journal}{Opt. Commun.} \textbf{\bibinfo{volume}{403}},
  \bibinfo{pages}{358} (\bibinfo{year}{2017}).

\end{thebibliography}

\end{document}